\documentclass[fleqn,usenatbib]{mnras}

\usepackage{newtxtext,newtxmath}

\usepackage[T1]{fontenc}
\usepackage{ae,aecompl}
\usepackage{gensymb}

\usepackage{graphicx}	
\usepackage{amsmath}	
\usepackage{amssymb}	

\newcommand{\objname}{NGTS-7Ab}
\newcommand{\starname}{NGTS-7A}
\newcommand{\bkgstarname}{NGTS-7B}
\newcommand{\numbername}{NGTS-7}
\newcommand{\startdate}{2016 May 04}
\newcommand{\stopdate}{2017 Jan 11}
\newcommand{\nightnumber}{130}
\newcommand{\periodhours}{16.2}

\newcommand{\minflareneergyone}{$7.7^{+2.4}_{-1.8}\times10^{32}$erg}
\newcommand{\maxflareenergyone}{$3.3^{+1.0}_{-0.8}\times10^{33}$erg}
\newcommand{\minflareneergytwo}{$2.5^{+0.7}_{-0.6}\times10^{32}$erg}
\newcommand{\maxflareenergytwo}{$1.1^{+0.3}_{-0.2}\times10^{32}$erg}
\newcommand{\flareoccrate}{$72\pm32$}
\newcommand{\vsini}{$v\sin i$}
\newcommand{\semiampvalue}{$25.9\pm0.9$~km~s$^{-1}$}
\newcommand{\midvaluepri}{$-4.2\pm0.8$~km~s$^{-1}$}
\newcommand{\midvaluebkg}{$-7.7\pm0.1$~km~s$^{-1}$}
\newcommand{\massvalueone}{$75.5^{+3.0}_{-13.7}$\,M$_\mathrm{J}$}
\newcommand{\massvaluetwo}{$48.5\pm4.3$\,M$_\mathrm{J}$}

\newcommand{\radiusone}{$1.38^{+0.13}_{-0.14}$\,R$_\mathrm{J}$}
\newcommand{\radiustwo}{$0.77\pm0.08$\,R$_\mathrm{J}$}

\newcommand{\neighbourage}{$55^{+80}_{-30}$}
\newcommand{\Gaia}{\textit{Gaia}}
\newcommand{\kepler}{\textit{Kepler}}
\newcommand{\corot}{\textit{CoRoT}}

\newcommand{\TESS}{\textit{TESS}}

\newcommand{\teff}{T$_{\mathrm{eff}}$}
\newcommand{\Rsun}{$R_{\odot}$}
\newcommand{\Msun}{$M_{\odot}$}

\newcommand{\mjup}{$M_\mathrm{J}$}

\newcommand{\lxlbol}{$\log L_{X}/L_{\mathrm{Bol}}$}
\newcommand{\lhlbol}{$\log L_{\mathrm{H{\alpha}}}/L_{\mathrm{Bol}}$}
\newcommand{\kms}{ km s$^{-1}$}
\newcommand{\kmsnospace}{km s$^{-1}$}
\newcommand{\ms}{ m s$^{-1}$}
\newcommand{\browndwarfnumber}{19}
\newcommand{\mstarbrowndwarfnumber}{4}

\newcommand{\logg}{$\log g$}
\newcommand{\emcee}{{\scshape emcee}}
\newcommand{\gillenbd}{AD 3116}
\newcommand{\rewrite}[1]{\textcolor{black}{#1}}
\newcommand{\ag}{$A_{\mathrm{g}}$}
\newcommand{\tbd}{$T_{\mathrm{BD}}$}

\title[\objname: An ultra-short period brown dwarf]{\objname: An ultra-short period brown dwarf transiting a tidally-locked and active M dwarf}
\author[J. A. G. Jackman et al.]{James A. G. Jackman,$^{1,2}$\thanks{E-mail: J.Jackman@warwick.ac.uk}
Peter J. Wheatley,$^{1,2}$\thanks{E-mail: P.J.Wheatley@warwick.ac.uk}
Dan Bayliss,$^{1,2}$
Samuel Gill,$^{1,2}$\newauthor
Simon~T.~Hodgkin,$^{3}$
Matthew R. Burleigh,$^{4}$
Ian P. Braker,$^{4}$\newauthor
Maximilian~N.~G{\"u}nther,$^{5,6,7}$ 
Tom Louden,$^{1,2}$
Oliver~Turner,$^{8}$
David~R.~Anderson,$^{1,2}$\newauthor
Claudia Belardi,$^{4}$
Fran\c{c}ois Bouchy,$^{8}$
Joshua T. Briegal,$^{5}$
Edward M. Bryant,$^{1,2}$\newauthor
Juan Cabrera,$^{9}$
Sarah L. Casewell,$^{4}$
Alexander Chaushev, $^{10}$
Jean C. Costes,$^{11}$\newauthor
Szilard Csizmadia,$^{9}$
Philipp Eigm\"uller, $^{9}$
Anders Erikson,$^{9}$
Boris T.~G\"ansicke,$^{1}$\newauthor
Edward Gillen,$^{5,\ddagger}$
Michael R. Goad,$^{4}$
James S. Jenkins,$^{12}$
James McCormac,$^{1,2}$\newauthor
Maximiliano Moyano,$^{14}$
Louise D. Nielsen,$^{9}$
Don Pollacco, $^{1,2}$\newauthor
Katja Poppenhaeger,$^{11,15}$
Didier Queloz, $^{5}$
Heike Rauer, $^{9,10,16}$
Liam Raynard,$^{4}$\newauthor
Alexis M. S. Smith,$^{9}$
St\'{e}phane Udry,$^{8}$
Jose I. Vines,$^{12}$
Christopher A. Watson,$^{11}$\newauthor
Richard~G.~West$^{1,2}$\\
Affiliations listed at the end of the paper}
\date{Accepted 2019 Sept 04. Received 2019 Aug 14; in original form 2019 Jun 19}

\pubyear{2019}

\begin{document}
\label{firstpage}
\pagerange{\pageref{firstpage}--\pageref{lastpage}}
\maketitle

\begin{abstract}
We present the discovery of \objname, a high mass brown dwarf transiting an M dwarf with a period of \periodhours\ hours, discovered as part of the Next Generation Transit Survey (NGTS). This is the shortest period transiting brown dwarf around a main or pre-main sequence star to date. The M star host (NGTS-7A) has an age of roughly 55 Myr and is in a state of spin-orbit synchronisation, which we attribute to tidal interaction with the brown dwarf acting to spin up the star. The host star is magnetically active and shows multiple flares across the NGTS and follow up lightcurves, which we use to probe the flare-starspot phase relation. The host star also has an M star companion at a separation of 1.13\,arcseconds with very similar proper motion and systemic velocity, suggesting the NGTS-7 system is a hierarchical triple. The combination of tidal synchronisation and magnetic braking is expected to drive ongoing decay of the brown dwarf orbit, with a remaining lifetime of only 5-10\,Myr. 
\end{abstract}
\section{Introduction} \label{sec:introduction}
The discovery of brown dwarfs in transiting exoplanet surveys provides a unique opportunity to probe the parameters of these substellar objects. With radii similar to Jupiter and masses between 13 and $\sim$78 \mjup\ \cite[e.g.][]{Chabrier00,Halbwachs00}, brown dwarfs are believed to form through molecular cloud fragmentation or gravitational instability, as opposed to the core accretion process that is commonly thought to form giant planets \citep[e.g.][]{Chabrier14}. Although the youngest substellar objects can have radii similar to early M stars \citep[e.g.][]{Stassun06}, as they age they undergo gravitational contraction \citep[][]{Lissauer04}. As brown dwarfs age their luminosity and temperature also decreases, resulting in their spectral energy distribution shifting towards longer 
wavelengths. As such, lone brown dwarfs can be identified in photometric surveys from their  
colours \citep[e.g.][]{Pinfield08,Folkes12,Rey18}. However, as close companions to pre-main or main sequence stars, such identification is not possible and we must rely on their effects on the host star.

The large masses of brown dwarfs should provide easily detectable signatures in radial velocity measurements relative to those of exoplanets \cite[e.g.\kms\ instead of\ms][]{Brahm16,Carmichael19}. Despite this, the number of transiting brown dwarfs relative to exoplanets remains low, with currently only \browndwarfnumber\ known to date.

The paucity of brown dwarfs on short periods around main sequence stars has previously been termed the ``brown dwarf desert'', from radial velocity and transit observations \citep[e.g.][]{Campbell98,Marcy00,Grether06}. This driving factor for this desert is typically attributed to the different formation mechanisms of low and high mass brown dwarfs in binary systems \citep[e.g.][]{Ma14}. High mass brown dwarfs ($\gtrapprox$ 43 \mjup) are believed to form through molecular cloud fragmentation, whereas their lower mass counterparts form within the protoplanetary disc. However, along with their formation pathways, a contributing element for the brown dwarf desert may be inward orbital migration of the brown dwarf \citep[e.g.][]{Armitage02}. One way of driving this is thought to be through tidal interactions between brown dwarfs and their host stars \citep[e.g.][]{Patzold02,Damiani16}, along with the effect of the magnetic braking of the host star \citep[e.g.][]{Barker09,Brown11}. If the companion is close enough, tidal interactions can decay its orbit, moving the companion inwards. The angular momentum lost from this orbit is expected to be transferred to the spin of the host star \citep[e.g.][]{Bolmont12}, eventually resulting in a state of spin-orbit synchronisation. In this state, the orbital and spin periods are equal. 
Such synchronisation has been detected in transiting brown dwarf systems before, for example in \corot-15b \citep[][]{Bouchy11}, a 63\mjup\ brown dwarf orbiting an F7V star with a period of 3.06 days. Along with this, brown dwarf systems have shown behaviour close to synchronisation \citep[e.g. WASP-128b][]{Hodzic18}. However, during this process, magnetic braking will remove angular momentum from the system \citep[][]{Barker09}. This acts to spin down the star, which in turn exacerbates the orbital decay of the companion. As such, even though (pseudo) spin-orbit synchronisation may be achieved, for active stars the magnetic braking can still drive the decay of the companion orbit. The combination of these effects eventually results in the engulfment of the brown dwarf by the host star. The timescale of this orbital decay is dependent on a number of factors, notably the stellar radius \citep[][]{Damiani16}. Consequently, the decay timescale is expected to be shortest for brown dwarfs around G and K type stars \citep[e.g.][]{Guillot14}, making brown dwarf companions rarer around these stars \citep[as noted by][]{Hodzic18} and contributing to the desert.

For M stars the orbital decay timescale is expected to be longer than G and K stars, due to the strong dependence of tides on stellar radius \citep[e.g.][]{Damiani16}. This is in spite of the strong magnetic activity of M stars, which can manifest itself as both saturated quiescent X-ray emission and transient activity such as stellar flares \citep[e.g.][]{Hilton11,Jackman18b}. Of the \browndwarfnumber\ transiting brown dwarfs  
known to date, 
\mstarbrowndwarfnumber\ brown dwarfs have been identified transiting M stars.
Two of these systems are hierarchical triples consisting of two M dwarfs and a brown dwarf \citep[NLTT41135 B, LHS 6343C;][]{Irwin11,Johnson11}, with the two M dwarfs in close proximity on the sky (2.4\arcsec, 55 AU and 0.55\arcsec, 20 AU respectively). Both these systems are believed to have ages greater than 1 Gyr and be in stable configurations.  
The third system, AD 3116 \citep[][]{Gillen17}, is a M+BD system discovered in the Praesepe open cluster and has an age of $\sim$ 700 Myr. This age makes it one of the younger transiting brown dwarf systems and useful for testing brown dwarf models with age. The fourth M+BD system is LP 261-75 \citep[][]{Irwin18}, a M+BD transiting pair with a distant visual brown dwarf companion \citep[][]{Reid06}. LP 261-75 is expected by \citet{Irwin18} to have an age of several Gyrs despite the high activity of the M dwarf primary, which instead suggests an age in the 130-200 Myr range \citep[e.g.][]{Reid06}. This strong activity instead is associated with tides from interactions between the brown dwarf and the host star. These four systems show the range of ages and configurations these systems can have, highlighting how further observations of transiting brown dwarfs are required to understand their formation and evolution. In particular, the discovery of unstable systems are needed in order to test evolutionary scenarios.

In this paper we report the discovery of \objname, a brown dwarf transiting an active M star on a 16.2 hour orbital period. The host star's rotation period is locked to the orbit of the brown dwarf, posing questions about the formation and evolution of such systems. We present our detection with NGTS, along with follow up photometric and spectroscopic measurements to constrain the radius and mass of the brown dwarf and M star host. We also present a detection of the secondary eclipse with NGTS, which we use to measure the temperature of \objname. This system is heavily diluted by a possibly associated nearby source. We describe the steps taken to account for this, along with presenting different scenarios based on the assumptions taken. We also discuss the possible formation scenarios of this system and outline how it may evolve in the future.

\section{Observations}\label{sec:observations}

\subsection{Photometry}
\subsubsection{NGTS} \label{sec:ngts}

\begin{table*}
	\centering
	\begin{tabular}{lccc}
    \hline
    Property & \starname & \bkgstarname & Source \tabularnewline \hline
    R.A [$\degree$] & 352.5216665551376 & 352.52202473338 &  1 \tabularnewline
    Dec [$\degree$]& -38.96992064512876 & -38.97006605140 & 1 \tabularnewline
    \Gaia\ Source ID & 6538398353024629888 & 6538398353024172032 & 1 \tabularnewline  
    $\mu_{R.A.}$ (mas yr$^{-1}$) & $-27.003\pm0.112$ & $-28.601\pm0.112$ & 1 \tabularnewline
    $\mu_{Dec}$ (mas yr$^{-1}$) & $-16.225\pm0.178$ & $-14.776\pm0.364$ & 1 \tabularnewline 
    Parallax (mas) & $7.2497\pm0.1203$ & $6.5232\pm0.0787$ & 1 \tabularnewline 
    $B$ & \multicolumn{2}{|c|}{$17.091\pm0.072$} & 2 \tabularnewline
    $V$ & \multicolumn{2}{|c|}{$15.502\pm0.028$} & 2\tabularnewline
    $g'$ & \multicolumn{2}{|c|}{$16.187\pm0.044$} & 2\tabularnewline
    $r'$ & \multicolumn{2}{|c|}{$14.940\pm0.010$} & 2\tabularnewline
    $i'$ & \multicolumn{2}{|c|}{$13.822\pm0.127$} & 2\tabularnewline
    \Gaia\ $G$ & $14.9154\pm0.0020$ & $15.5134\pm0.0012$ & 1 \tabularnewline
    $J$ & \multicolumn{2}{|c|}{$11.832\pm0.030$} & 3 \tabularnewline
    $H$ & \multicolumn{2}{|c|}{$11.145\pm0.026$} & 3 \tabularnewline
    $K_{s}$ & \multicolumn{2}{|c|}{$10.870\pm0.019$} & 3 \tabularnewline
    $W1$ & \multicolumn{2}{|c|}{$10.740\pm0.022$} & 4 \tabularnewline
    $W2$ & \multicolumn{2}{|c|}{$10.660\pm0.020$} & 4 \tabularnewline
	\hline
	\end{tabular}
	\caption{Stellar properties for each star. We have listed the photometry used in our SED fitting. We show the parallax and proper motions for reference, however do not use them all in our analysis for the reasons outlined in Sect.\,\ref{sec:astrometry}. The references are: 1. \citet{Gaia18}, 2. \citet{APASS_14}, 3. \citet{2MASS_2006}, 4. \citet{ALLWISE2014}. }
	\label{tab:stellar_params}
\end{table*}
\begin{figure}
	\includegraphics[width=\columnwidth]{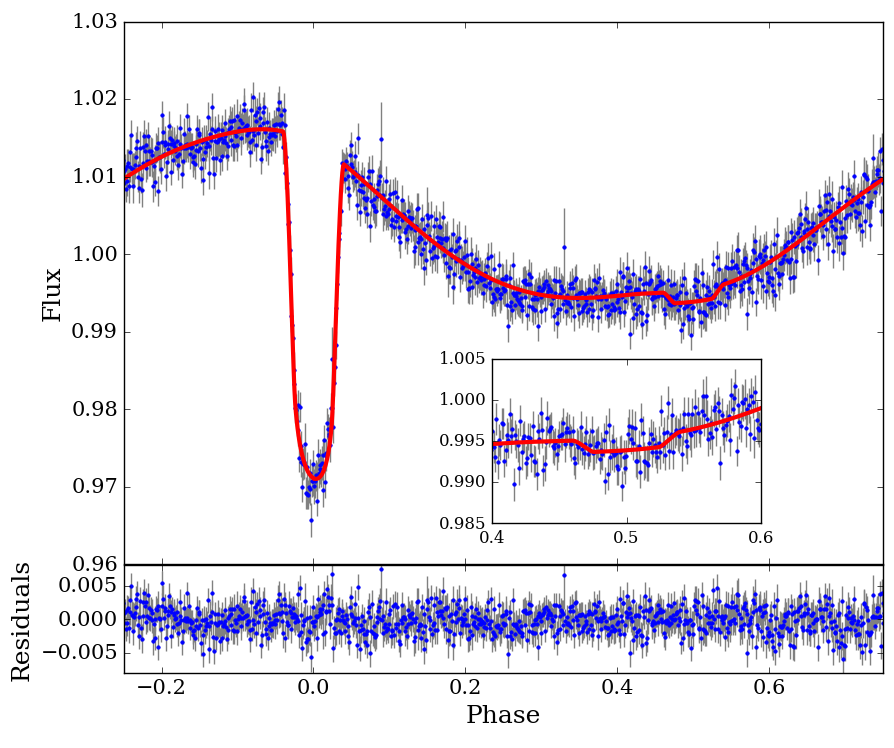}
    \caption{The binned, phase-folded NGTS lightcurve showing both the transit and starspot modulation. The NGTS data (in blue) has been placed into 1000 bins, equal to approximately 1 minute each. We have overlaid the best fitting model in red. The inset plot shows a zoom in of the secondary eclipse. Lower panel shows the residuals of our fitting.}
    \label{fig:transit_plot}
\end{figure}
\numbername\ was observed with NGTS for \nightnumber\ nights between \startdate\ and \stopdate, using a single camera. The phase folded lightcurve is shown in Fig.\,\ref{fig:transit_plot}. Observations were obtained in the custom NGTS filter (520-890nm) with a cadence of 13 seconds. For a full description of the NGTS instrument and pipeline processing see \citet{Wheatley18}. The NGTS lightcurves were detrended using a version of the {\scshape sysrem} algorithm, as done for previous NGTS discoveries \citep[e.g.][]{Bayliss17,Raynard18,West18}. 

This star was originally identified as an object of interest due to the detection of flares as part of the NGTS flare survey \citep[e.g.][]{Jackman18a,Jackman18b}. We subsequently identified a \periodhours\ hour periodicity. We then noted transit events of 4.3 per cent depth occurring on the same period.

\Gaia\ DR2 resolves two stars with a separation of 1.13\arcsec, while all other catalogues list it as a single source. \rewrite{The catalogue photometry and astrometry is given in Tab.\,\ref{tab:stellar_params}. To confirm the source of the transits we perform centroiding using the vetting procedure described by \citet{Gunther17}. We describe this analysis in Sect.\,\ref{sec:transit_confirm} and refer to the two sources as \starname\ and \bkgstarname\, where \starname\ is the transit source. The two stars have \Gaia\ G magnitudes of 14.9 (\starname) and 15.5 (\bkgstarname), meaning that there is non-negligible dilution present in our photometry, something we discuss and account for in Sect.\,\ref{sec:fitting}.}

\subsubsection{SAAO}
\begin{figure}
	\includegraphics[width=\columnwidth]{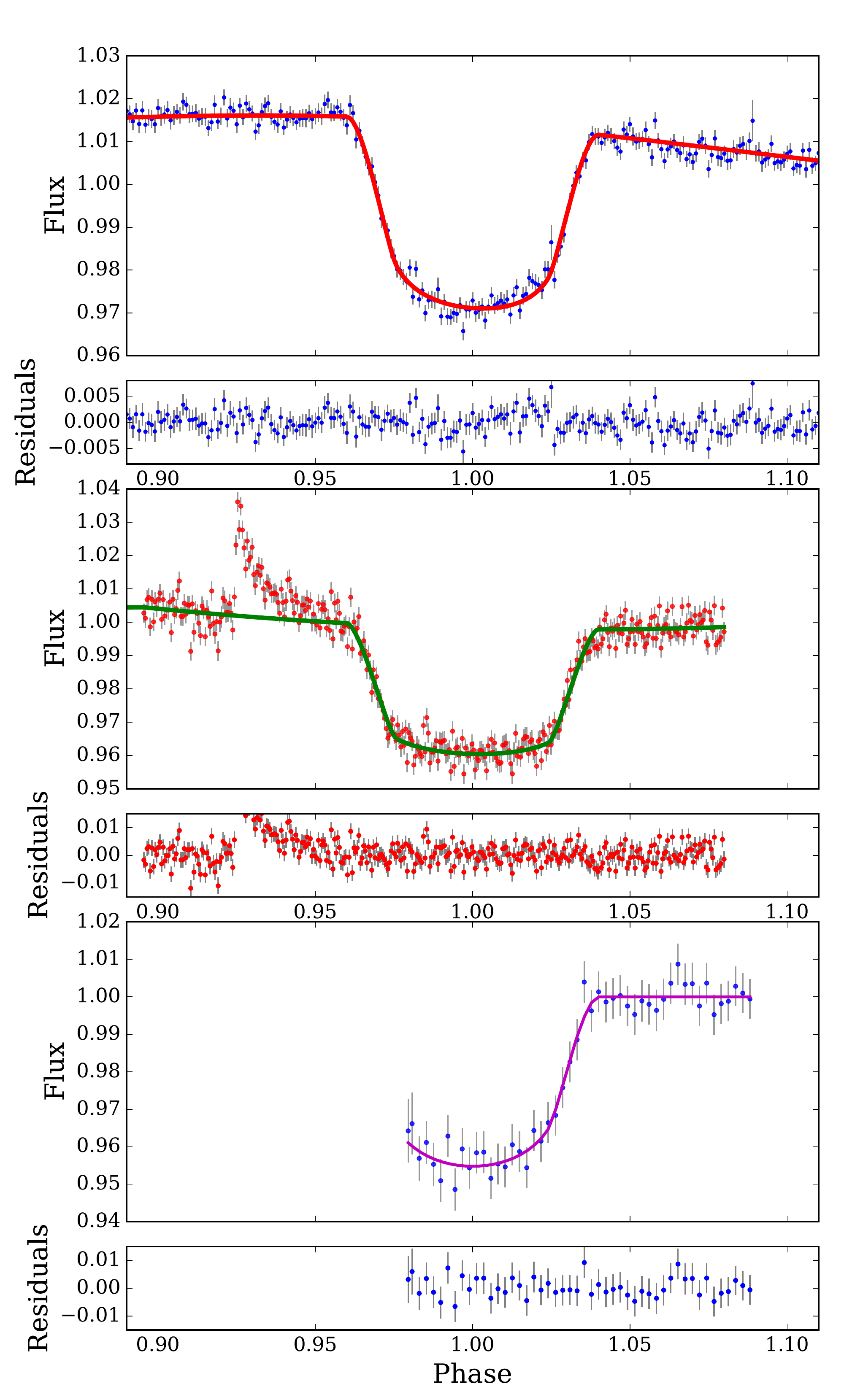}
    \caption{Transit lightcurves of \objname. Top: phase folded NGTS lightcurve (as in Fig.\,\ref{fig:transit_plot}) with the best fitting model overlaid in red. Middle: Primary transit lightcurve from SAAO in I band, with the best fitting model in green. Bottom: Primary transit lightcurve from EulerCam in V band, with the best fitting model in magenta. Residuals for each fit are shown underneath each respective plot.}
    \label{fig:follow_up}
\end{figure}

Follow up photometry of \numbername\  
was obtained at the South African Astronomical Observatory (SAAO) on 2018 Aug 08 (\textit{I} band, secondary eclipse), 2018 Aug 11 (\textit{I} band, primary transit), and again on 2018 Oct 04 (\textit{I} band, secondary eclipse) using the 1.0m Elizabeth telescope and ``{\it shocnawe}'', one of the SHOC high speed CCD cameras \citep[][]{Coppejans13}. On each occasion, sky conditions were clear throughout the observations, with the seeing around 2~arcseconds. 
The data were reduced with the local SAAO SHOC pipeline developed by Marissa Kotze, which is driven by {\sc python} scripts running {\sc iraf} tasks ({\sc pyfits} and {\sc pyraf}), and incorporating the usual bias and flat-field calibrations. Aperture photometry was performed using the {\sc Starlink} package {\sc autophotom}. We used a 5~pixel radius aperture that maximised the signal to noise ratio, and the background was measured in an annulus surrounding this aperture. One bright comparison star in the $2.85\times2.85$~arcminute field of view was then used to perform differential photometry on the target. The two stars identified by \Gaia\ DR2 coincident with the position of \numbername\ were not resolved in these data. Figure \ref{fig:follow_up} shows the primary transit observed on 2018 Aug 11. A stellar flare can be clearly seen shortly before transit ingress. 

\subsubsection{EulerCam}
One transit of \numbername\ 
was observed with EulerCam on the 1.2m Euler Telescope at La Silla Observatory \citep{Lendl12}. These observations were obtained on the night of 2018 Sept 01, in the \textit{V} band filter and are shown in Fig.\,\ref{fig:follow_up}. The data were bias and flat field corrected then reduced using the PyRAF implementation of the ``{\scshape phot}" routine. An aperture radius and ensemble of comparison stars were used such that the scatter in the out of transit portion of the lightcurve was minimised. 

\subsubsection{TESS}

\begin{figure*}
\includegraphics[width=\textwidth]{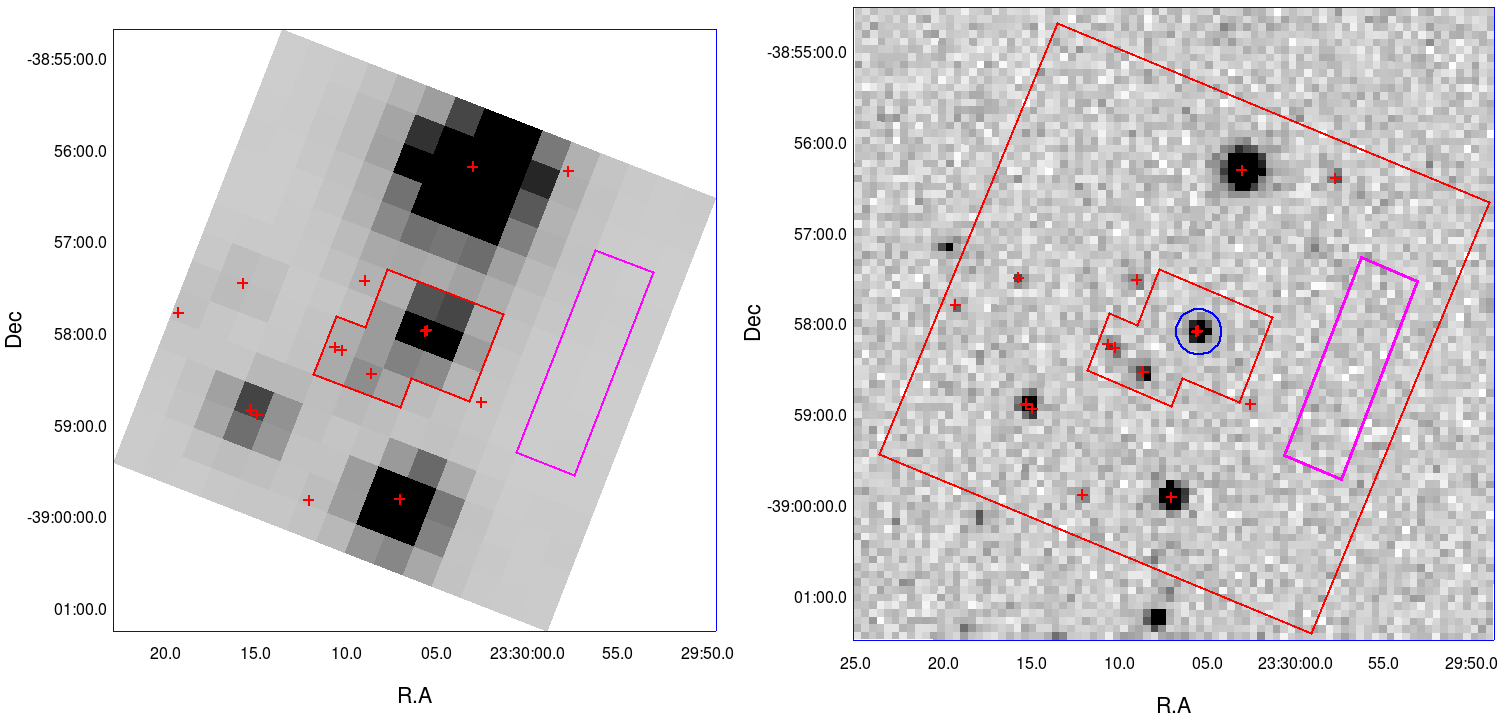}
\caption{Left: the first image of the \TESS\ full-frame image stack (15 $\times$ 15 pixel) showing the field surrounding \starname. Nearby companions with \Gaia\ magnitudes brighter than $G = 18.4$ (3.5 magnitudes fainter than \starname) are plotted with red crosses. The aperture used to extract the \TESS\ lightcurve is outlined in red. We subtracted the per-pixel background contribution estimated from selecting a region (outlined in magenta) free of \Gaia\ stars brighter than $G = 18.4$. Right: an example NGTS image of the same region of sky with the TESS FFI region shown. The aperture used for the NGTS photometry is shown in blue. We have overlaid the \TESS\ apertures in this image for reference.} 
    \label{fig:103323_field}
\end{figure*}

\begin{figure}
	\includegraphics[width=\columnwidth]{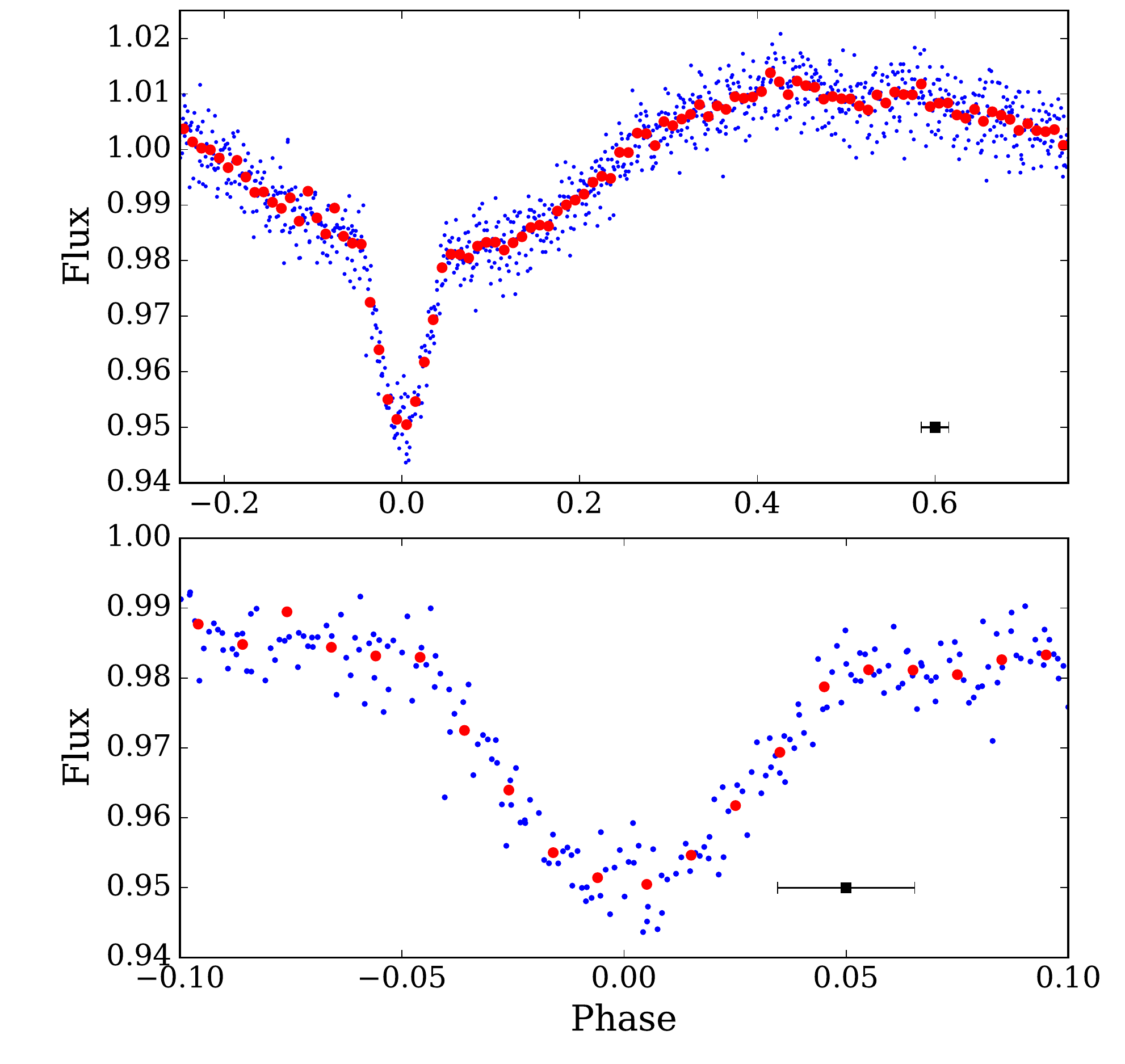}
    \caption{Top: Phase folded \TESS\ lightcurve from the sector 2 full frame images. Blue points indicate individual \TESS\ observations and the red points are the phase fold binned to 100 bins in phase. The black point represents the length of an individual 30 minute cadence \TESS\ observation in phase. Note how the out of transit modulation has changed in phase from the original NGTS observations. Bottom: Zoom in of the primary transit. Note the more V-shaped appearance of the primary transit compared to those in Fig.\,\ref{fig:follow_up}, due to the smearing effect of the 30 minute cadence observations.}
    \label{fig:TESS_lc}
\end{figure}

\numbername\ was observed at a 30 minute cadence with the NASA Transiting Exoplanet Survey Satellite (\TESS) 
\citep[][]{Ricker15} between 2018 Aug 27 and 2018 Sep 19, in Sector 2. A 15 $\times$ 15 pixel (5.25' $\times$ 5.25') cutout was obtained from the \TESS\ full-frame image stacks using the \textsc{tesscut} routine \footnote{https://github.com/spacetelescope/tesscut}. This cutout is shown in Fig. \ref{fig:103323_field}. Aperture masks were chosen by-eye to exclude nearby bright sources up to 2.5' away. The 21" pixel-scale of \TESS\ creates a PSF of 
\numbername\ which is blended with at least 3 significantly bright stars ($\Delta G < 3.5$ mag.) As it is not possible to completely exclude the flux from these blended stars in \TESS\ we chose our aperture to enclose them, with the knowledge the \TESS\ lightcurve will be diluted. We estimated the per-pixel background contribution by selecting 8 pixels West of the aperture that do not include any stars brighter than $G = 18.4$ (3.5 magnitudes fainter than \starname). This region is shown as the magenta box in Fig.\,\ref{fig:103323_field}. This was subtracted from the aperture-summed flux to create a background-corrected light curve. 

The transit seen in the \TESS\ light curve is both shallower and more V-shaped than that from NGTS, despite the similar bandpasses of NGTS and \TESS. This is due to a combination of additional dilution in the \TESS\ data (from the neighbouring sources) and the 30 min cadence which smears out the transit \citep[which has a duration of only 1.3 hours, e.g.][]{Smith18}. Due to these effects we do not use the \TESS\ lightcurve in our transit fitting (Sect.\,\ref{sec:fitting}). \rewrite{However, we do use it in Sect.\,\ref{sec:starspots} where we discuss the phase of the out-of-transit variations of \numbername.}

\subsection{Spectroscopy}
\subsubsection{HARPS} \label{sec:HARPSS}
\begin{figure}
	\includegraphics[width=\columnwidth]{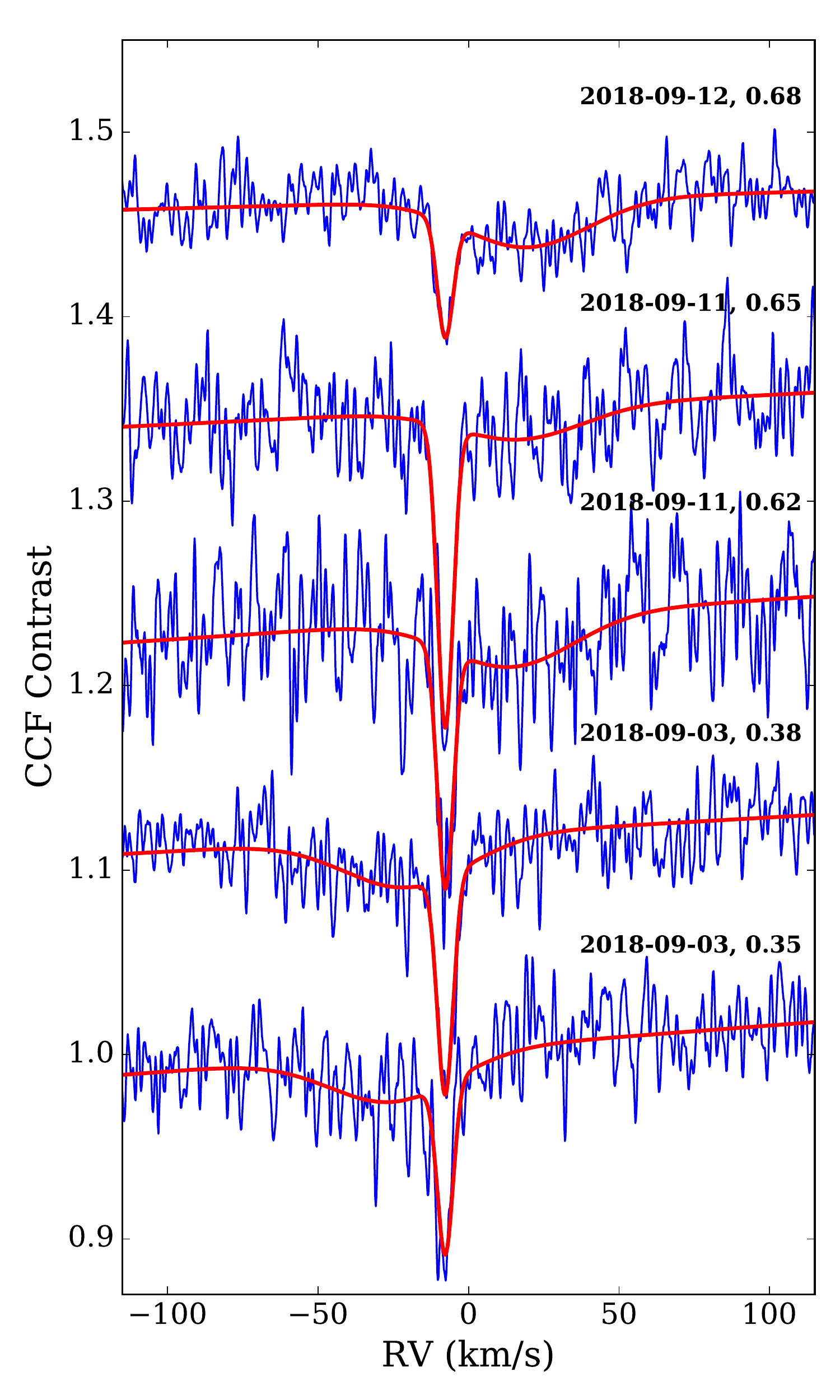}
    \caption{CCFs from HARPS, using a K5 mask and offset in contrast. The HARPS data is shown blue, with observation date and orbital phase for each CCF provided. For each HARPS CCF we have simultaneously fitted two Gaussians 
    along with a varying baseline, which are overlaid in red. We can see that along with the narrow peak with a constant RV due to \bkgstarname, there is a clear shift of a wide Gaussian, which we attribute to \starname.}
    \label{fig:ccf_plot}
\end{figure}
We obtained high-resolution spectroscopy for \starname\ with the HARPS spectrograph on the ESO 3.6m telescope \citep[][]{Mayor03}. Five measurements \rewrite{with an exposure time of 1800s} were taken on the nights beginning 2018 Sept 02 and 2018 Sept 11 as part of programme ID 0101.C-0889(A). Due to the relative faintness 
of the source we used the high efficiency fibre link (EGGS), with 
a fibre size of 1.4\arcsec\ instead of the usual 1.0\arcsec\ mode. Consequently, these spectra contain light from both \starname\ and \bkgstarname\ \rewrite{and we see a narrow and a broad peak in the Cross Correlation Functions (CCFs) shown in Fig.\,\ref{fig:ccf_plot}}. \rewrite{The RVs of \starname\ and \bkgstarname\ along with the respective contrasts from our analysis in Sect.\,\ref{sec:radial_velocity} are given in Tab.\,\ref{tab:HARPS_rvs}.}

\subsubsection{SAAO} \label{sec:spectrum}
\begin{figure*}
	\includegraphics[width=\textwidth]{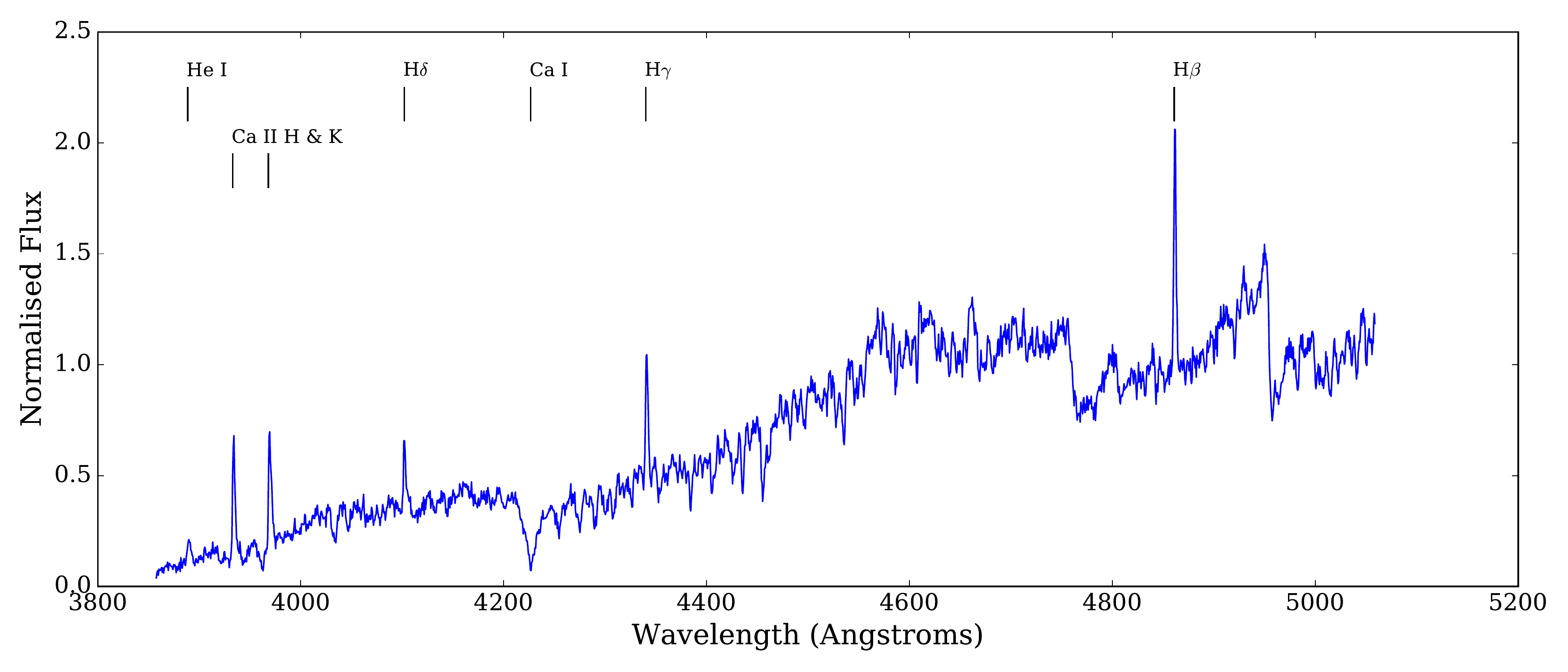}
    \caption{SAAO spectrum of NGTS YA+B \starname\ with H, He and Ca emission and absorption lines marked. The spectrum has been normalised to the flux at 5000\,\AA. The emission lines show \starname\ is chromospherically active.}
    \label{fig:spectrum}
\end{figure*}

Follow up spectroscopy of \numbername\ was also obtained from SAAO on the 1.9m telescope using the SpUpNIC instrument \citep[][]{Spupnic16} between the dates 2018 Sept 09 and 2018 Sept 11. 14 spectra with a resolution of R=2500 were obtained in total, with a wavelength range of 3860-5060\AA. We have combined these spectra to give the average spectrum shown in Fig.\,\ref{fig:spectrum}. Observations were performed with a slit width of 1.8\arcsec\ and average seeing of 2\arcsec, once again meaning both \starname\ and \bkgstarname\ are present in our data. Figure \ref{fig:spectrum} shows clear TiO and CaI absorption features expected for M dwarf spectra. Along with this, we see several emission lines from the Balmer series, as well as He I and Ca II, showing at least one of the stars is chromospherically active. 

\section{Analysis}
The observations of Sect.\,\ref{sec:observations} were combined with available catalogue photometric and astrometric information. We use this information to \rewrite{confirm the source of the transits and  }characterise both \starname\ and \bkgstarname.

\subsection{Identifying the Source of the Transit} \label{sec:transit_confirm}

\begin{figure*}
\begin{minipage}{0.26\linewidth}
\vspace*{-7.0cm}
\includegraphics[width=\textwidth, height=\textwidth, angle=0]{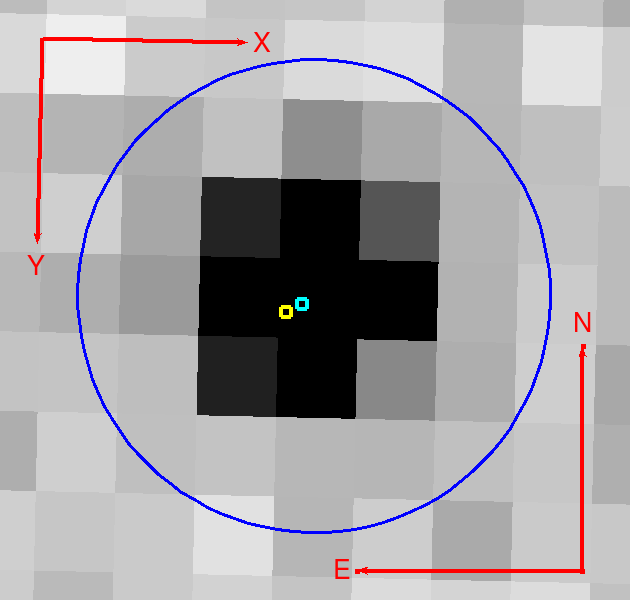}
\end{minipage}\qquad
\includegraphics[width=0.55\linewidth]{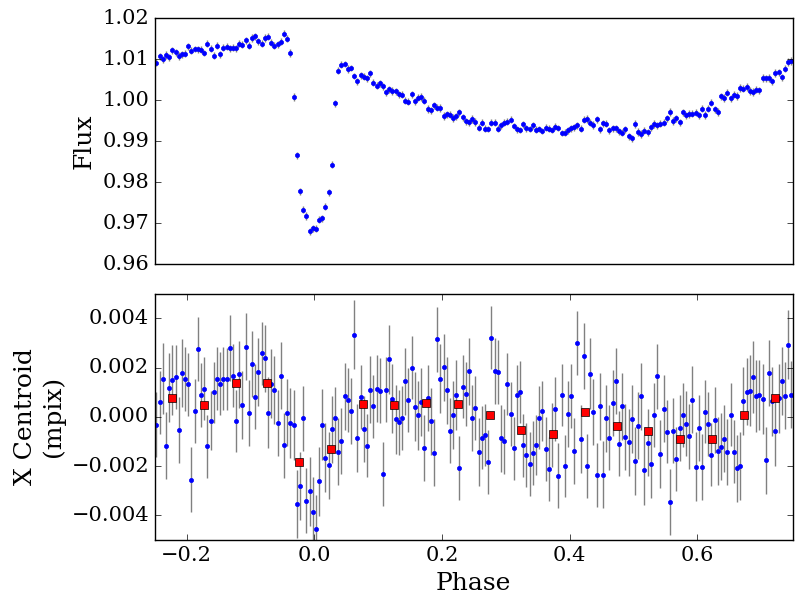}

\caption{\rewrite{Left: NGTS image of \numbername\ with the aperture shown as the blue circle. The \Gaia\ DR2 positions of \starname\ and \bkgstarname\ are shown with the cyan and yellow points respectively. Right: Our centroid analysis of \numbername. The top panel shows the phase folded NGTS lightcurve, placed into 200 bins. The bottom panel shows the phase folded X centroid position. The blue points indicate the same 200 bins as the top panel, while the red points are the same data binned up by a factor of 10. We can see both the clear centroid movement during the transit and with the out of transit modulation.}}
\label{fig:aperture}

\end{figure*}

\rewrite{In Sect.\,\ref{sec:observations} we noted that \Gaia\ DR2 resolves two sources with a separation of 1.13\arcsec\ at the position of \numbername. To confirm which source our transit signal is coming from,  we performed centroiding using the vetting procedure described by \citet{Gunther17}. We identify that the transit and the out-of-transit modulation comes from \Gaia\ DR2 6538398353024629888, the brighter of the two sources. Figure \ref{fig:aperture} shows the phase folded transit and X centroid position, showing how the shape of the phase-folded centroid data follows the shape of the phase-folded lightcurve. While individual NGTS pixels are 5'' across, the NGTS centroiding procedure is able to identify centroid shifts below 1'' in size, meaning we are confident that we have identified the correct host star and now refer to this star as the primary star, or \starname. We refer to the neighbouring source as \bkgstarname\ and discuss it further in Sect.\,\ref{sec:wide_binary}.}

\rewrite{Out of transit modulation on the orbital period can be due to either ellipsoidal variation \citep[e.g.][]{Drake03,Welsh10} or reflection effects \citep[e.g.][]{Armstrong16,Eigmuller18}. However, neither of these could adequately explain the number or position in phase of the maxima seen in Fig.\,\ref{fig:transit_plot} (just before the primary transit). The most natural explanation is that this out-of-transit modulation is due to starspots on the host star and that the spin period of \starname\ is locked to the orbital period of the transiting body. This places \starname\ in a state of spin-orbit synchronisation \citep[e.g.][]{Ogilvie14}. The change in the out of transit modulation in the \TESS\ data can be explained by the evolution of starspots in the interval between the NGTS and \TESS\ observations.}

\rewrite{The 16.2 hour period rotation of \starname\ will result in its observed CCF in our HARPS spectra being rotationally broadened. This broadened peak will also move around with a 16.2 hour period. In Sect.\,\ref{sec:HARPSS} we noted that our HARPS spectra contain light from \starname\ and \bkgstarname\ and the presence of a narrow and broad peak in our HARPS CCFs, seen in Fig.\,\ref{fig:ccf_plot}. As we will discuss in Sect.\,\ref{sec:radial_velocity} we find the broad peak moves on a 16.2 hour period, as we might expect if \starname\ has a transiting body and itself is in a state of spin-orbit synchronisation. The rapid rotation of \starname\ and it being chromospherically active (as evidenced by the observed starspots) presumably means \starname\ is the source of the multiple stellar flares in the NGTS and SAAO lightcurves \citep[e.g.][]{Hawley14}. Along with this, \starname\ is likely the dominant source of the observed emission lines in our SAAO spectra.}

\rewrite{Based on our observations and the evidence presented here we are confident that \starname\ is the source of the observed transits. Along with this we believe \starname\ is in a state of spin-orbit synchronisation with its companion, which will have spun up \starname\ to keep it at the observed period.}

\subsection{Stellar Parameters} \label{sec:SED}
Throughout this paper, all of our 
photometry is measured in apertures which contain the light from both \starname\ and \bkgstarname. Consequently, in order to obtain accurate parameters for \objname\ we need to estimate the dilution from \bkgstarname. We have done this through fitting the Spectral Energy Distribution (SED) of both \starname\ and \bkgstarname, using a combination of the information available from blended catalogue photometry and \Gaia\ photometric and astrometric data. This information is listed in Tab.\,\ref{tab:stellar_params}. 

\subsubsection{\Gaia\ Photometry} \label{sec:gaia_phot}
\begin{figure}
\includegraphics[width=\columnwidth]{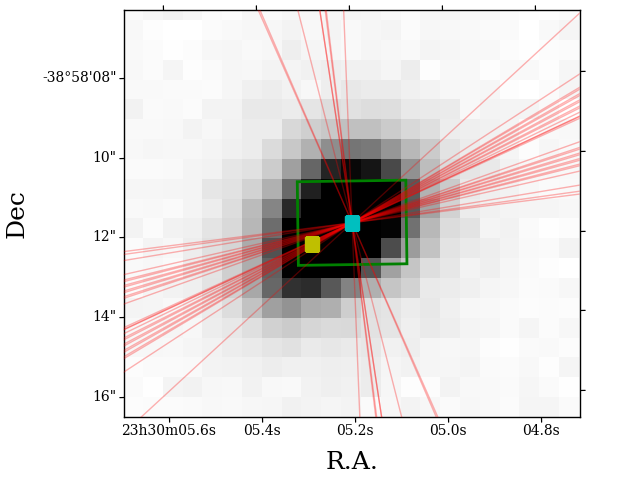}
\caption{SkyMapper i band image of \starname\ and \bkgstarname, with their \Gaia\ DR2 positions shown as the cyan and yellow squares respectively. The green rectangle is an example 3.5x2.1 arcsec$^{2}$ region used for calculating the \textit{BP} and \textit{RP} flux, centered on \starname. The red lines are the \Gaia\ scan directions obtained from GOST, which we have fixed to pass through the centre of \starname. Note the significant fraction that pass through, or close to, both \starname\ and \bkgstarname.}
    \label{fig:gost_image}
\end{figure}

While both sources have a \Gaia\ \textit{G} magnitude, only the primary star has \textit{BP} and \textit{RP} photometry. The \Gaia\ \textit{G} magnitudes for both stars are published in \Gaia\ DR2, and are derived from fitting the line spread function (LSF) of each star from windows which are approximately 0.7$\times$2.1 arcsec$^{2}$ in the along scan and across scan directions respectively \citep[][]{Gaia16}. 
We used the \Gaia\ Observation Scheduling Tool (GOST\footnote{https://gaia.esac.esa.int/gost/}) to check the scans of \starname\ and \bkgstarname\ used for \Gaia\ DR2. We obtained 35 scans, which are plotted over a SkyMapper \textit{i} band image \citep[][]{Wolf18} in Fig.\,\ref{fig:gost_image}. From Fig.\,\ref{fig:gost_image} we can see that over 75 per cent of the scans which went into \Gaia\ DR2 fall along (or close to) the position angle separating the two stars. \rewrite{Given the 1.13\arcsec\ separation of the stars and the ability of the LSF to resolve sources of this separation \citep[e.g. Fig. 7 in][]{Fabricius16}, we expect only minimal contamination between the stars in the \Gaia\ \textit{G} band photometry. Consequently, we use the \Gaia\ \textit{G} band photometry in our analysis.}

On the other hand, the \textit{BP} and \textit{RP} photometry is measured from the total flux in a 3.5x2.1 arcsec$^{2}$ region \citep[][]{Evans18}. An example of this region is shown in Fig.\,\ref{fig:gost_image}, showing that the \textit{BP} and \textit{RP} photometry will be of both \starname\ and \bkgstarname\ combined. This is reflected in the {\scshape \textit{BP}-\textit{RP} excess factor} of 2.054 for \starname. The {\scshape \textit{BP}-\textit{RP} excess factor} is the sum of light from the \textit{BP} and \textit{RP} bands compared to the \textit{G} band, and should ideally be around 1 for a single, non-contaminated, star. A value around 2 suggests the \textit{BP} and \textit{RP} photometry is 
comprised of flux from two similar stars.
Indeed, we find that the \textit{BP} and \textit{RP} photometry of \starname\ fails the filter from \citet{Arenou18}, which is used to remove contaminated stars from their analysis. Consequently we do not use the \Gaia\ \textit{BP} and \textit{RP} photometry of \starname\ in our analysis.

\subsubsection{\Gaia\ Astrometry} \label{sec:astrometry}
For both sources we initially test the quality of the \Gaia\ astrometry by calculating both the Unit Weight Error (UWE) and the Renormalised UWE (RUWE). We compare the UWE against the filter specified by \citet{Lindegren18} and check whether the RUWE is below the recommended value of 1.4 for a clean astrometric sample. We found that \starname\ suffers from significant astrometric excess noise ({\scshape astrometic excess noise sig}=71.6, RUWE=3.4), resulting in it failing both filters. \bkgstarname, while having non-zero astrometric excess noise ({\scshape astrometic excess noise sig}=4.4, RUWE=1.3), passes both filters. When calculating the astrometric solution of each star, \Gaia\ DR2 assumes a single object. 
The astrometric excess noise is the extra noise that is required by the single source solution to fit the observed behaviour. High levels of astrometric excess noise \rewrite{are} a sign that the single source solution has failed, possibly due to unresolved binarity \citep[e.g.][]{GaiaHR}. We also check each star further by comparing them against sources of similar magnitude, colour and parallax in the full \Gaia\ DR2 sample. Both stars are outliers from the main sample in terms of their astrometric quality. We note in particular that each has a correlation between their parallax and proper motion components. One possibility for the low quality of the astrometric parameters for \starname\ and \bkgstarname\ may be levels of blending due to their proximity. \citet{Lindegren18} has noted that during scanning of close sources the components can become confused, through a changing photocentre. 

\rewrite{Due to it failing the recommended astrometry filters we have decided not to use the astrometric solution of \starname\ in our analysis. As we explain in Sect.\,\ref{sec:wide_binary} we consider two scenarios. The first of these uses only the astrometric solution of \bkgstarname\ and 
fixes both \starname\ and \bkgstarname\ at the distance of \bkgstarname, while the second doesn't use \Gaia\ parallaxes and assumes both sources are on the main sequence.}

\subsubsection{A possible wide binary} \label{sec:wide_binary}
A scenario mentioned in Sect.\,\ref{sec:astrometry} which may be responsible for the low quality astrometry of 
\starname\ and \bkgstarname\ is that the two sources are a wide binary. If they are a wide binary, then we would expect them to be at the same distance.
\starname\ and \bkgstarname\ have very similar proper motions, which 
supports this assumption, which are shown for reference in Tab.\,\ref{tab:stellar_params}. However, as the proper motions are measured as part of the \Gaia\ astrometry and may have levels of contamination, we have decided to seek out additional evidence. Wide binaries have previously been identified in both TGAS \citep[e.g.][]{Andrews17} and \Gaia\ DR2, with \citet{Andrews18} finding that, as expected, real binaries will have similar systemic velocities, whereas chance alignments will not. From our radial velocity analysis in \rewrite{Sect.\,\ref{sec:radial_velocity}} we found that \starname\ and \bkgstarname\ have systemic velocities of \midvaluepri\ and \midvaluebkg\ respectively. Using the distance of \bkgstarname\ results in a projected separation of 173 AU. This projected separation and the difference in systemic velocities places \starname\ and \bkgstarname\ well within the \citet{Andrews18} sample of genuine wide binaries, instead of being a chance alignment on the sky.  Consequently, it is very likely that \starname\ and \bkgstarname\  
are in fact a wide binary and are 
at the same distance. If so, this would provide a way of constraining the distance to \objname, along with placing it in a hierarchical triple system. \rewrite{Checking for possible memberships of known associations using the BANYAN $\Sigma$ online tool \footnote{http://www.exoplanetes.umontreal.ca/banyan/banyansigma.php} reveals no likely associations \citep[][]{Gagne18}}.

Following this, we have devised two separate scenarios on the assumption that \starname\ and \bkgstarname\ are in a wide binary. These are as follows:
\begin{enumerate}
  \item We fix both sources at the distance of \bkgstarname, 
  assuming they are a wide binary.
  \item We believe neither \Gaia\ DR2 parallax, instead fixing them at the same distance and assuming they are on the main sequence.
\end{enumerate}
These scenarios both avoid using the poor astrometric solution of \starname.

\subsubsection{SED Fitting}
To determine the SED of both stars we have fitted two separate components simultaneously using a custom SED fitting process which utilises the PHOENIX v2 grid of models \citep[][]{Husser13}, following a similar method to \citet{Gillen17}. Initially we generated a grid of bandpass fluxes and spectra in \teff-\logg\ space, which allowed us to interpolate across these parameters. 
We fit for \teff, \logg, along with the radius, $R$, and distance, $D$, of each star. We have chosen to fix the metallicity at the Solar value. 
Prior to fitting we inflated the errors of catalogue photometry by 2.5 per cent 
to account for the observed variability in the NGTS lightcurve. During fitting we compare the combination of fluxes from each star to the observed values, for all filters in Tab.\,\ref{tab:stellar_params} except \Gaia\ G (which is used as a prior to normalise the respective SEDs). 
To explore the full posterior parameter space we use \emcee\ \citep[][]{emcee} to generate an MCMC process, using 200 walkers for 50,000 steps, disregarding the first 25,000 as a burn in. 

We have used a range of physically motivated priors in our modelling which we outline here. Firstly, the radii and distances are used in our model to scale the flux from each star by $(R/D)^{2}$. For scenario (i) (Sect.\,\ref{sec:wide_binary}) we have placed a Gaussian prior on the distance of each star, using the value from \cite{BailerJones18} for \bkgstarname, $152.7\pm1.9$ pc. In this scenario the fitted radius of each star is allowed to vary freely. For scenario (ii) we fit for the distance, which we also force to be the same for the two stars. 
We have placed a Gaussian prior on the fitted radius for each star, using the \citet{Mann15} \teff-radius relation. For this prior, we have used the 13.4 per cent error given by \citet{Mann15} as the standard deviation of the Gaussian prior 
to allow some variation. In both scenarios we have placed a prior on the synthetic \Gaia\ G band flux for each star, using the observed flux values. This was done to anchor each star to observations.

\begin{figure}
	\includegraphics[width=\columnwidth]{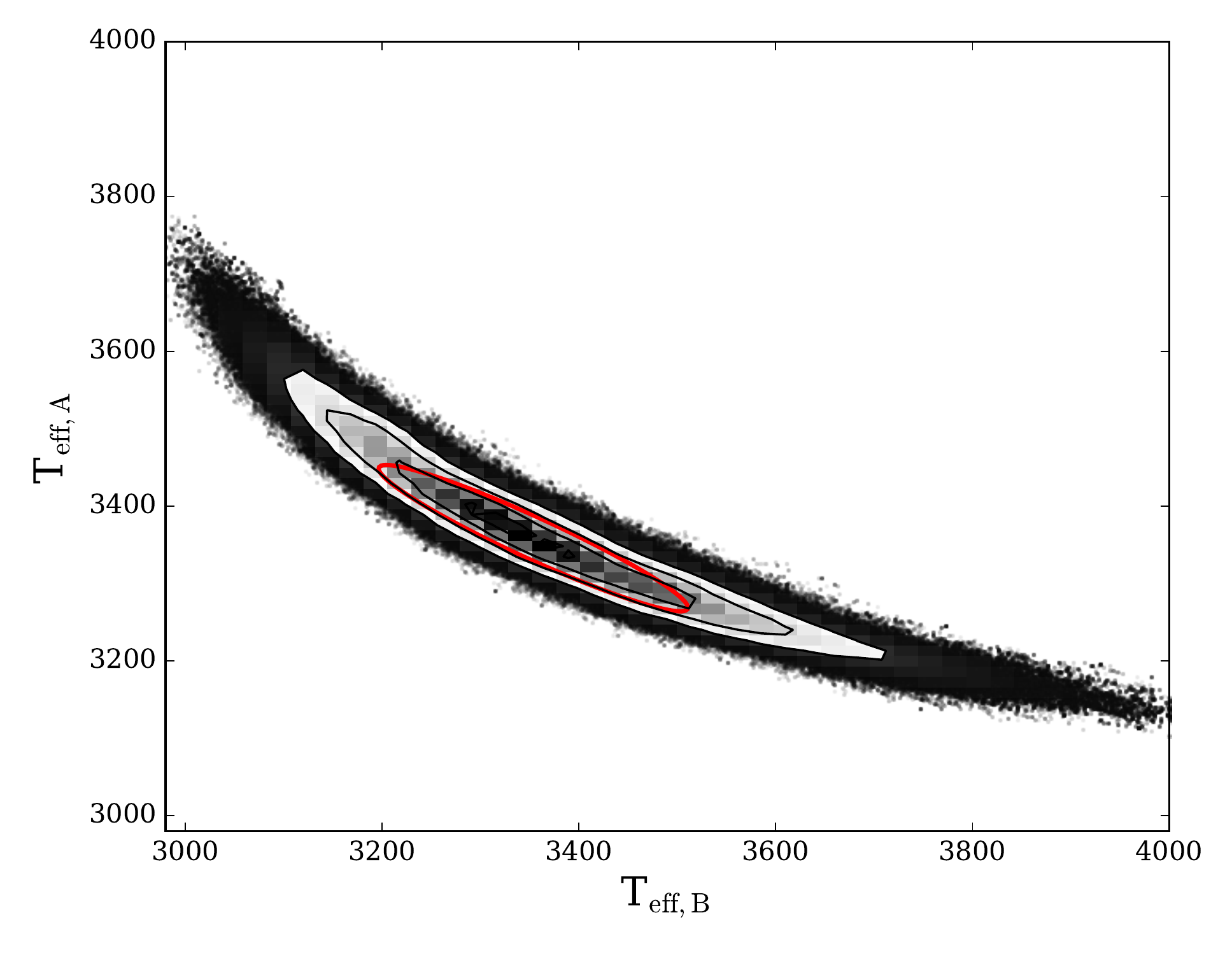}
    \caption{The posterior \teff$\mathrm{_{,A}}$-\teff$\mathrm{_{,B}}$ distribution of the scenario (i) (Sect.\,\ref{sec:wide_binary}) SED fit, showing the correlation between the effective temperatures of \starname\ and \bkgstarname. The red ellipse indicates the estimate of the 1$\sigma$ region.}
    \label{fig:corner_teff}
\end{figure}

Table \ref{tab:fitting_params_table} gives the results of each fit. For both scenarios  
we retrieve two stars with temperatures 
corresponding to M3-M4 spectral type \citep[e.g.][]{Pecaut13}. This similarity in spectral type matches what we would expect from the \Gaia\ BP-RP excess factor, as discussed in Sect.\,\ref{sec:gaia_phot}. However, for scenarios (i) and (ii) 
we measure very 
different stellar radii for \starname\ and \bkgstarname. Investigating the posterior distribution of our SED fit reveals a strong correlation between the \teff\ and radius values of \starname\ and \bkgstarname. An example of this for \teff\ is shown in Fig.\,\ref{fig:corner_teff}, with the full corner plot shown in Fig.\,\ref{fig:corner_fig}. The full corner plot for scenario (ii) is shown in Fig.\,\ref{fig:corner_fig_main_seq}. 
The correlation between \teff\ and radius arises from the similarity of the two sources in spectral type, along with the availability of only 
the \Gaia\ G magnitude to separate them. 
This correlation needs to be taken into account when determining the uncertainties in the age and mass of \starname. To incorporate these correlations we fit the 2D posterior distributions from our SED fitting with ellipses covering 68 per cent of our distribution. 
We have used these ellipses to probe the extremes of parameter space and incorporate the observed correlations into our analysis (Sect.\,\ref{sec:primary_mass}). For each parameter we also report the 16th, 50th and 84th percentile of the marginalised 1D distributions in Tab.\,\ref{tab:fitting_params_table}.

\subsubsection{Primary Mass} \label{sec:primary_mass}
When we fix both stars to the \Gaia\ distance for \bkgstarname\ (scenario (i)), 
the median radius of the primary star is approximately 75 per cent oversized in radius compared to that of 
a main sequence star of the median \teff. One possible reason for this is that \starname\ and \bkgstarname\ 
are pre-main sequence stars and as such both have a larger than expected radius \citep[e.g][]{Jackman18b}. 
In order to estimate the mass of the primary star we compared each source to the PARSEC isochrones \citep[][]{Bressan12}, assuming that \starname\ and \bkgstarname\ are the same age (reasonable if we assume they are bound). From comparing the median radius and \teff\ of \bkgstarname\ to the PARSEC isochrones we obtained an age estimate of 55 Myr and a mass of 0.35\,\Msun\ for \bkgstarname. However, using this age estimate with the fitted parameters of \starname\ results in different mass estimates based on whether we use the median \teff\ (0.35 \Msun) or the radius (0.55 \Msun).

A potential reason for this discrepancy is the effect of starspots on \starname. For both main and pre-main sequence stars, modelling of starspots has shown they can act to both increase the stellar radius and decrease \teff\,\citep[][]{Jackson14,Somers16}. 
The combined effect of these changes can be a diminished stellar luminosity \citep[][]{Jackson14}, which results in discrepancies when comparing to unspotted stellar models. 

\begin{figure}
	\includegraphics[width=\columnwidth]{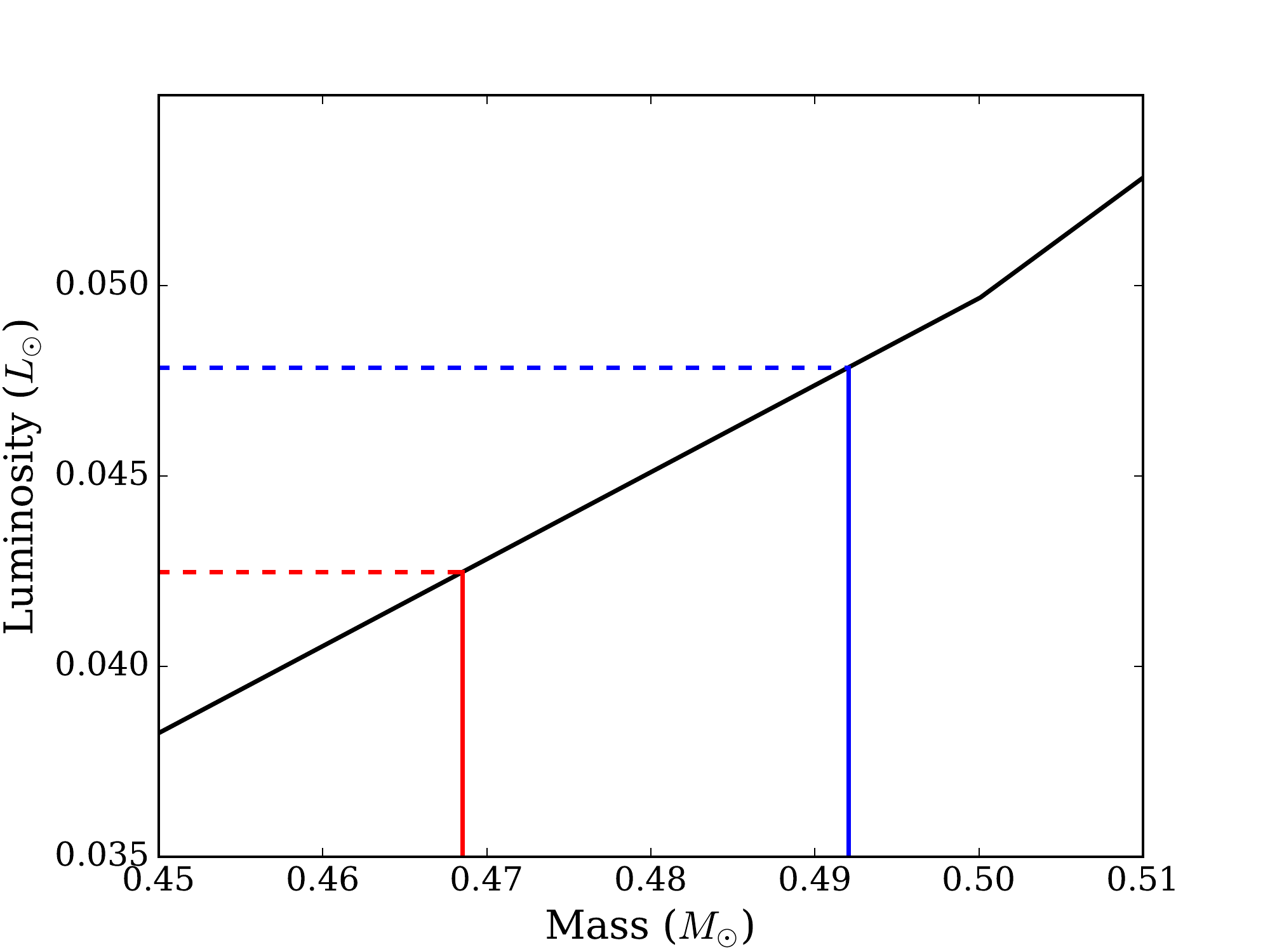}
    \caption{An example of the starspot correction described in Sect.\,\ref{sec:SED}. In black is the PARSEC mass-luminosity isochrone for 55 Myr. The red and blue lines correspond the unspotted models with luminosity equal to our SED fit and before the appearance of spots respectively.}
    \label{fig:luminosity_correction}
\end{figure}
To correct for the effects of spots on our mass estimate for a given age, we used the PARSEC models to identify which unspotted models give a luminosity equal to or up to 10 per cent greater than the current value \citep[this is approximately the change in luminosity caused by the sudden appearance of spots simulated by][]{Jackson14}. This was done for the median \teff\ and radius values of each parameter, as given in Tab.\,\ref{tab:stellar_params}. In this analysis we used the 1$\sigma$ extremes determined from the posterior distribution error ellipses from Sect.\,\ref{sec:SED}, in order to take correlations into account. This resulted in an age of \neighbourage\ Myr for \bkgstarname. We calculate the luminosity of \starname\ based on the median and the error ellipse also, to give a range of possible masses depending on the obtained primary parameters and age of \bkgstarname. Using the possible ages of the neighbour to determine the unspotted model we estimate the mass of the primary as  $0.48^{+0.03}_{-0.12}$\Msun. For the errors we have combined the extremes from the age of the neighbour and whether the luminosity is altered by the appearance of spots. 

Based on the age estimate of \neighbourage\ Myr for this system we have also searched for signs of Li 6708\AA\, absorption in our HARPS spectra. Primordial lithium is quickly depleted within the interiors of M stars \citep[e.g.][]{Chabrier96} and is typically removed from their photosphere within 45-50 Myr \citep[see Fig. 4 of][]{Murphy18}. We do not find any sign of Li 6708\AA\, absorption in our HARPS spectra, consistent with our estimate of \neighbourage\ Myr and it suggests that the system cannot be much younger than this if scenario (i) is true.

We note we have assumed in this section that the companion star does not also suffer from spots, which may be unlikely for a young system. The presence of spots would alter the inferred age and hence mass of the primary star. However, we do not identify any significant modulation in either the NGTS or \TESS\ lightcurves which could be attributed to spots on the companion. 

We also note that while there exist empirical relations to attempt to correct for the effects of magnetic activity on measured \teff\ and radii \citep{Stassun12}, using the ratio of the H$\alpha$ and bolometric luminosity, \lhlbol. These relations are used to bring the \teff\ and radius values closer to expected model values, which can then be used to calculate the age and mass of \starname. 
Unfortunately it is likely that our measurements of H$\alpha$ luminosity for \starname\ are contaminated by \bkgstarname\ to an uncertain degree (from H$\alpha$ emission of its own). Consequently, we have chosen not to use these relations to adjust our fitted values here, but do discuss this further in Sect.\,\ref{sec:issues}.

For the second scenario where we have assumed both stars are drawn from the \citet{Mann15} \teff-radius relation, we calculate a distance of $88.04^{+8.91}_{-8.79}$pc, given in Tab.\ref{tab:fitting_params_table}. To calculate the mass in this scenario we use the empirical mass relation of \cite{Benedict16} for main sequence M stars. We have calculated the value of $M_\mathrm{K_{s}}$ for \starname\ using the best fitting SED model and the fitted distance. Using this relation with calculated distance of $88.04^{+8.91}_{-8.79}$pc for the \starname\ we calculate the primary mass $M_\mathrm{A}$ to be $0.24\pm0.03$\Msun.

\begin{figure}
	\includegraphics[width=\columnwidth]{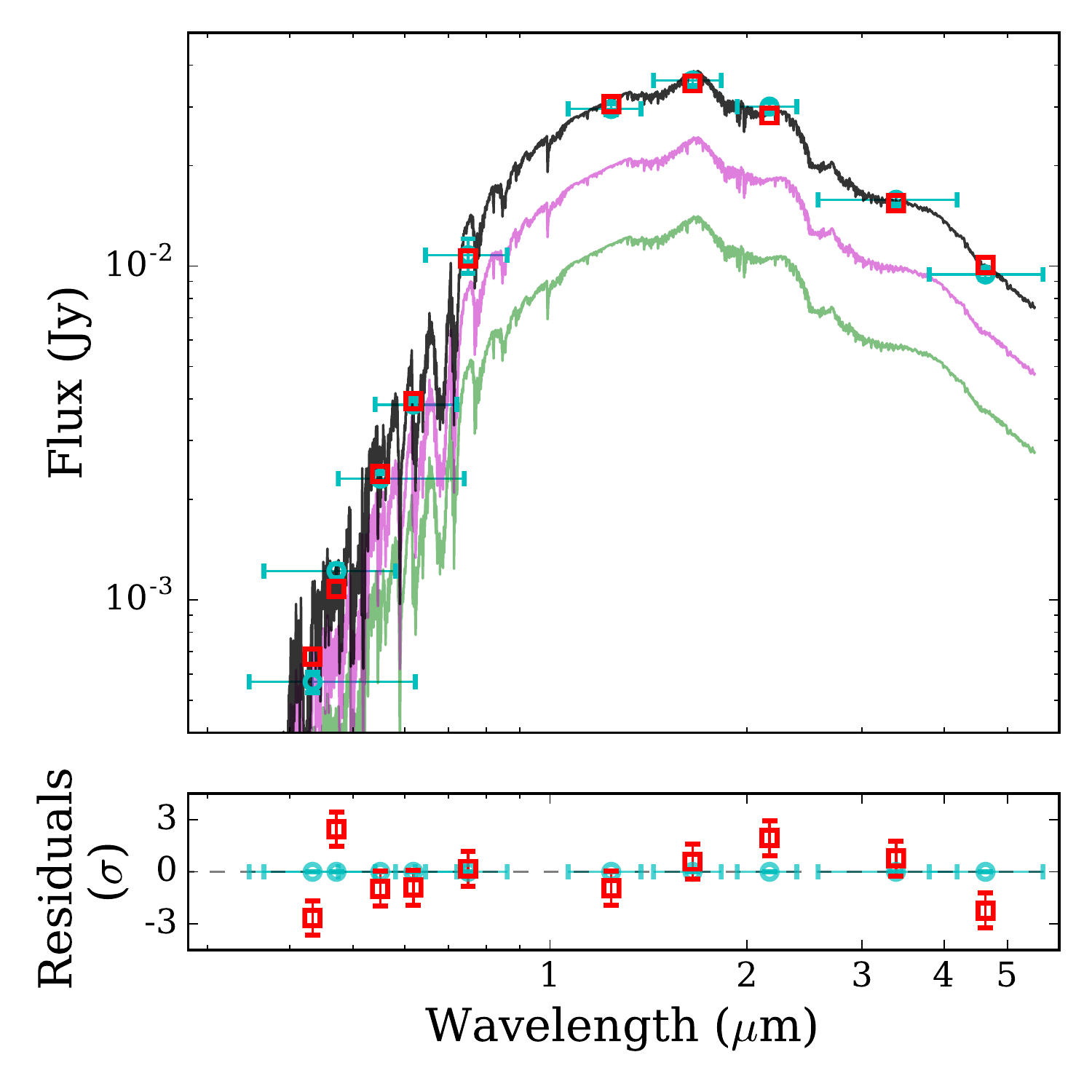}
    \caption{Top: The best fitting two-component PHOENIX v2 SED model for scenario (i). The magenta and green curves are the best fitting models for \starname\ and \bkgstarname, while the cyan and red points indicate the catalogue and synthetic photometry respectively. The horizontal error bars represent the spectral range of each filter. Bottom: Residuals of the synthetic photometry, normalised to the adjusted catalogue errors.}
    \label{fig:SED_fit}
\end{figure}

\renewcommand{\arraystretch}{1.25}

\begin{table*}
	\centering
	\caption{Parameters from our fitting of NGTS-7AB system for the scenarios defined in Sect.\,\ref{sec:wide_binary}. In scenario (i) we have placed both stars at the distance of \bkgstarname\ and fit for the radius, whereas in scenario (ii) we assumed both stars were on the main sequence and fit for both radius and distance. The bold values for scenario (i) are to indicate it is our favoured scenario, as discussed in Sect.\,\ref{sec:nature}. Here we report the median of the 1D distribution for each parameter, along with the errors determined from the 16th and 84th percentiles. Limb darkening parameters with asterisks had priors applied when fitting (Sect.\,\ref{sec:fitting}).}
	\label{tab:fitting_params_table}
	\begin{tabular}{lcc}
	Scenario & \textbf{(i)} & (ii) \tabularnewline
    \hline
    SED Fitting \tabularnewline \hline
    \teff$_{A}$ (K) & $\mathbf{3359^{+106}_{-89}}$& $3393^{+30}_{-31}$\tabularnewline
    \teff$_{B}$ (K) & $\mathbf{3354^{+172}_{-147}}$ & $3300^{+44}_{-42}$
 \tabularnewline
    \logg$_{A}$ & $\mathbf{4.89^{+0.40}_{-0.28}}$
& $4.82^{+0.39}_{-0.23}$ \tabularnewline
    \logg$_{B}$ & $\mathbf{4.98^{+0.37}_{-0.34}}$ & $4.99^{+0.36}_{-0.34}$ \tabularnewline
    $R_{A}$ (\Rsun) & $\mathbf{0.61^{+0.06}_{-0.06}}$ & $0.34^{+0.04}_{-0.04}$ \tabularnewline
    $R_{B}$ (\Rsun) & $\mathbf{0.46^{+0.08}_{-0.07}}$ & $0.28^{+0.03}_{-0.03}$ \tabularnewline
    $D_{A}$ (pc)  & $\mathbf{152.67^{+2.01}_{-2.01}}$& $88.04^{+8.91}_{-8.79}$ \tabularnewline
    $D_{B}$ (pc)  & $\mathbf{152.70^{+1.98}_{-1.99}}$ & $88.04^{+8.91}_{-8.79}$ \tabularnewline \hline

    Transit Parameters \tabularnewline \hline
    Period (hours) & $\mathbf{16.2237952^{+0.0000026}_{-0.0000018}}$ & 16.2237957$^{+0.0000024}_{-0.0000021}$ \tabularnewline
    Time of transit centre (days) $T_{centre}$ (HJD - 2456658.5) &
    $\mathbf{1050.053304^{+0.0000017}_{-0.0000055}}$ & 1050.053311$^{+0.0000099}_{-0.0000125}$ \tabularnewline
    $R_{A}/a$ & $\mathbf{0.20213^{+0.00310}_{-0.00257}}$ & $0.20215^{+0.00366}_{-0.00258}$ \tabularnewline
    $R_{BD}/a$ & $\mathbf{0.04710^{+0.00093}_{-0.00061}}$ & $0.04725^{+0.00121}_{-0.00062}$ \tabularnewline
    $a$ (AU) & $\mathbf{0.0139^{+0.0013}_{-0.0014}}$ & $0.0078^{+0.0009}_{-0.0008}$  \tabularnewline
    $i$ (\degree) & $\mathbf{88.43520^{+0.98314}_{-1.10843}}$ & $88.43124^{+1.01065}_{-1.29644}$ \tabularnewline
    Surface brightness ratio & $\mathbf{0.03620^{+0.01148}_{-0.01198}}$ & $0.03763^{+0.01296}_{-0.01225}$ \tabularnewline
    SAAO LD1* & $\mathbf{0.24872^{+0.02043}_{-0.02002}}$  & $0.25023^{+0.02006}_{-0.02080}$ \tabularnewline
    SAAO LD2 & $\mathbf{0.06045^{+0.12719}_{-0.12362}}$ & $0.06297^{+0.12759}_{-0.14502}$ \tabularnewline
    EulerCam LD1* & $\mathbf{0.53550^{+0.01645}_{-0.01732}}$ & $0.53480^{+0.01760}_{-0.01690}$ \tabularnewline
    EulerCam LD2 & $\mathbf{0.15415^{+0.19985}_{-0.24625}}$ & $0.17269^{+0.17810}_{-0.22253}$ \tabularnewline
    NGTS LD1* & $\mathbf{0.36273^{+0.02752}_{-0.05013}}$ & $0.36208^{+0.02798}_{-0.05178}$ \tabularnewline
    NGTS LD2 & $\mathbf{0.38254^{+0.12664}_{-0.11373}}$ & $0.36759^{+0.14530}_{-0.12993}$ \tabularnewline \hline
    Spot Parameters & & \tabularnewline \hline
    Spot 1 $l$ ($\degree$) & $\mathbf{74.68895^{+3.82344}_{-3.21393}}$ & $75.22438^{+4.62037}_{-3.48933}$ \tabularnewline
    Spot 1 $b$ ($\degree$)  & $\mathbf{50.01602^{+8.70891}_{-11.89659}}$ & $49.48639^{+9.24656}_{-12.49710}$ \tabularnewline
    Spot 1 size ($\degree$)  & $\mathbf{13.87737^{+3.60749}_{-2.67395}}$ & $13.76827^{+3.05474}_{-2.56000}$ \tabularnewline
    Spot 1 brightness factor & $\mathbf{0.48236^{+0.18490}_{-0.25854}}$ & $0.46430^{+0.17512}_{-0.24309}$ \tabularnewline
    Spot 2 $l$ ($\degree$)  & $\mathbf{176.06974^{+4.63372}_{-3.76279}}$ & $176.58879^{+5.62414}_{-4.08154}$ \tabularnewline
    Spot 2 $b$ ($\degree$)  & $\mathbf{77.97929^{+1.81747}_{-2.10508}}$ & $77.47726^{+1.98424}_{-2.56784}$ \tabularnewline
    Spot 2 size ($\degree$)  & $\mathbf{30.25273^{+3.62982}_{-3.82615}}$  & $30.22503^{+3.72268}_{-4.27825}$ \tabularnewline
    Spot 2 brightness factor & $\mathbf{0.27168^{+0.16630}_{-0.17487}}$  & $0.30954^{+0.15558}_{-0.19153}$ \tabularnewline \hline
	\end{tabular}
\end{table*}

\renewcommand{\arraystretch}{1}

\subsection{Transit and Spot Fitting} \label{sec:fitting}
In order to model the transits of \objname\ we used the ELLC package \citep[][]{Maxted16}. ELLC is a binary star model that 
allows for multiple spots to be included on each star and as such can be used to model both transits and spot modulation at the same time. 

We simultaneously fit the NGTS, SAAO and EulerCam lightcurves to ensure consistent transit parameters across our entire dataset. For the NGTS data we fit a transit model combined with a two spot model, to account for the out of transit modulation. We tested our fitting using both a single and double spot model, however we found a single spot was unable to match the average out of transit behaviour seen in Fig.\,\ref{fig:transit_plot}. The transit in the \TESS\ data is blurred by the 30\,min cadence of the observations, and also suffers additional dilution from a number of blended sources (see Fig.\ref{fig:103323_field}), and so we decided not to include the \TESS\ light curve in our fit. 
We can use the \TESS\ data to see spot modulation has changed between the NGTS observations and the SAAO and EulerCam follow up lightcurves (which were obtained at similar times to the \TESS\ data). Consequently we did not use the NGTS spot model to fit the SAAO or EulerCam follow up lightcurves. However, the SAAO lightcurve of primary transit on 2018 Aug 11 does show evidence of the spot minimum during the single night, consistent with the \TESS\ data. We incorporated this into our fitting as a quadratic term which we fit simultaneously with the transits. 

The SAAO lightcurve also includes a flare just before ingress, which we masked out for our fitting but analyse in Sect.\,\ref{sec:magnetic_activity}. 
For each bandpass we directly fitted independent limb darkening profiles. We used a quadratic limb darkening profile and generated our initial limb darkening parameters using the Limb Darkening Toolkit, \citep[LDtk;][]{ldtk}, using the best fitting SED from Sect.\,\ref{sec:SED}. During fitting we allowed each second limb darkening coefficient (LD2) to vary, while keeping the first (LD1) constant to reduce degeneracy in the fit. For each photometric band we also incorporate a dilution term, to account for the flux from the neighbouring star. For each band we use a Gaussian prior 
based on the expected dilution (and standard deviation) from our SED fitting. To estimate the expected dilution in a given bandpass we convolve the SED for each star with the specified filter curve and take the ratio of measured values. In order to take the observed correlations into account we sample the expected values for the Gaussian prior directly from the posterior distribution of the SED fits. For each filter we use the dilution term to correct the transit model as
\begin{equation}
    \delta_{filter} = \bigg(\frac{R_{BD}}{R_{A}}\bigg)^{2} \bigg(1 + \bigg(\frac{F_{B}}{F_{A}}\bigg)\bigg)^{-1}
\end{equation}
where $\delta_{filter}$, $R_{BD}$, $R_{A}$ are the transit depth in the chosen filter, radii of the companion and \starname\ respectively, while $F_{B}$ and $F_{A}$ are the fluxes of \bkgstarname\ and \starname\ in the specified bandpass. In the ideal scenario where $F_{B}$=0 we can see this becomes the usual transit depth equation. During our preliminary fitting we found the eccentricity to be consistent with zero when applying the \citet{Lucy71} criterion. 
Consequently for our final fitting we fixed the eccentricity at zero, i.e that the orbit has circularised. Due to the high time cadence of NGTS it is not feasible to fit the entire NGTS lightcurve for each step of the MCMC process. Instead we bin the lightcurve to 1000 bins in phase, using the period and epoch specified for that step. We chose 1000 bins in order to preserve the information in the ingress and egress. In order to sample the posterior parameter space we used \emcee\ with 200 walkers for 50,000 steps and disregarding the first 25,000 as a burn in. We did this for both scenarios (i) and (ii), using the dilution values from the relevant SED model. The values of the best fitting parameters are shown in Tab.\,\ref{tab:fitting_params_table}.

Using the results of our transit fitting for scenarios (i) and (ii), we measure the radius of \objname\ to be \radiusone\ and \radiustwo\ respectively. As brown dwarfs are expected to shrink with age \citep[e.g.][]{Baraffe03}, scenario (i) would imply a younger brown dwarf than scenario (ii), consistent with our age estimation from Sect.\,\ref{sec:primary_mass}. The single period in our fitting is able to model both the orbital and spin periods, supporting our conclusion in Sect.\,\ref{sec:transit_confirm} that the system is in a state of spin-orbit synchronisation. 

Our best fitting spot model suggests the presence of two spot regions with a large size and a low brightness factor. Each region can be interpreted either as a single large spot of constant brightness, or as a series of smaller, darker, spots spread over a similar area. 
As we only fit for the dominant spots, our model is unable to rule out the presence of spots elsewhere on the star. It is most likely that these are smaller than our fitted regions however, as large spots elsewhere could act to decrease the observed variability \citep[e.g.][]{Rackham18}.

\subsection{Radial Velocity} \label{sec:radial_velocity}
\newcommand{\specialcell}[2][l]{%
  \begin{tabular}[#1]{@{}c@{}}#2\end{tabular}}

\begin{table*}
	\centering
	\begin{tabular}{lccccccc}
    \hline
    \specialcell{$\mathrm{BJD_{\,TDB}}$\\(-2,450,000)} & \specialcell{$\mathrm{RV_{A}}$\\(\kmsnospace)} & \specialcell{$\mathrm{RV_{A}}$ error\\(\kmsnospace)}&\specialcell{$\mathrm{Contrast_{A}}$\\(per cent)} & \specialcell{$\mathrm{RV_{B}}$\\(\kmsnospace)} & \specialcell{$\mathrm{RV_{B}}$ error\\(\kmsnospace)} & \specialcell{$\mathrm{Contrast_{B}}$\\(per cent)} & S/N\tabularnewline \hline
    8364.50765417 & -25.369 & 1.513 & 2.584 & -7.751 & 0.054 & 9.261 & 6.1\tabularnewline
    8364.52902662 & -20.596 & 1.640 & 2.673 & -7.751 & 0.054 & 11.787& 6.7 \tabularnewline
    8373.47768215 & 14.435 & 2.326 & 2.703 &-7.751 & 0.054 & 12.936& 4.5 \tabularnewline
    8373.49931132 & 17.142 & 3.578 & 1.744 & -7.751 & 0.054 & 16.270 & 5.8 \tabularnewline
    8373.52064613 & 19.385 & 1.159 & 2.608 & -7.751 & 0.054 & 6.286 & 7.7 \tabularnewline

	\hline
	\end{tabular}
	\caption{\rewrite{HARPS radial velocities for \starname\ and \bkgstarname\ from our analysis in Sect.\,\ref{sec:radial_velocity}. The radial velocity of \bkgstarname\ is fixed to be constant during our analysis. The signal-to-noise ratios correspond to the spectral order 66 centered at 653 nm.}}
	\label{tab:HARPS_rvs}
\end{table*}

\begin{figure}
	\includegraphics[width=\columnwidth]{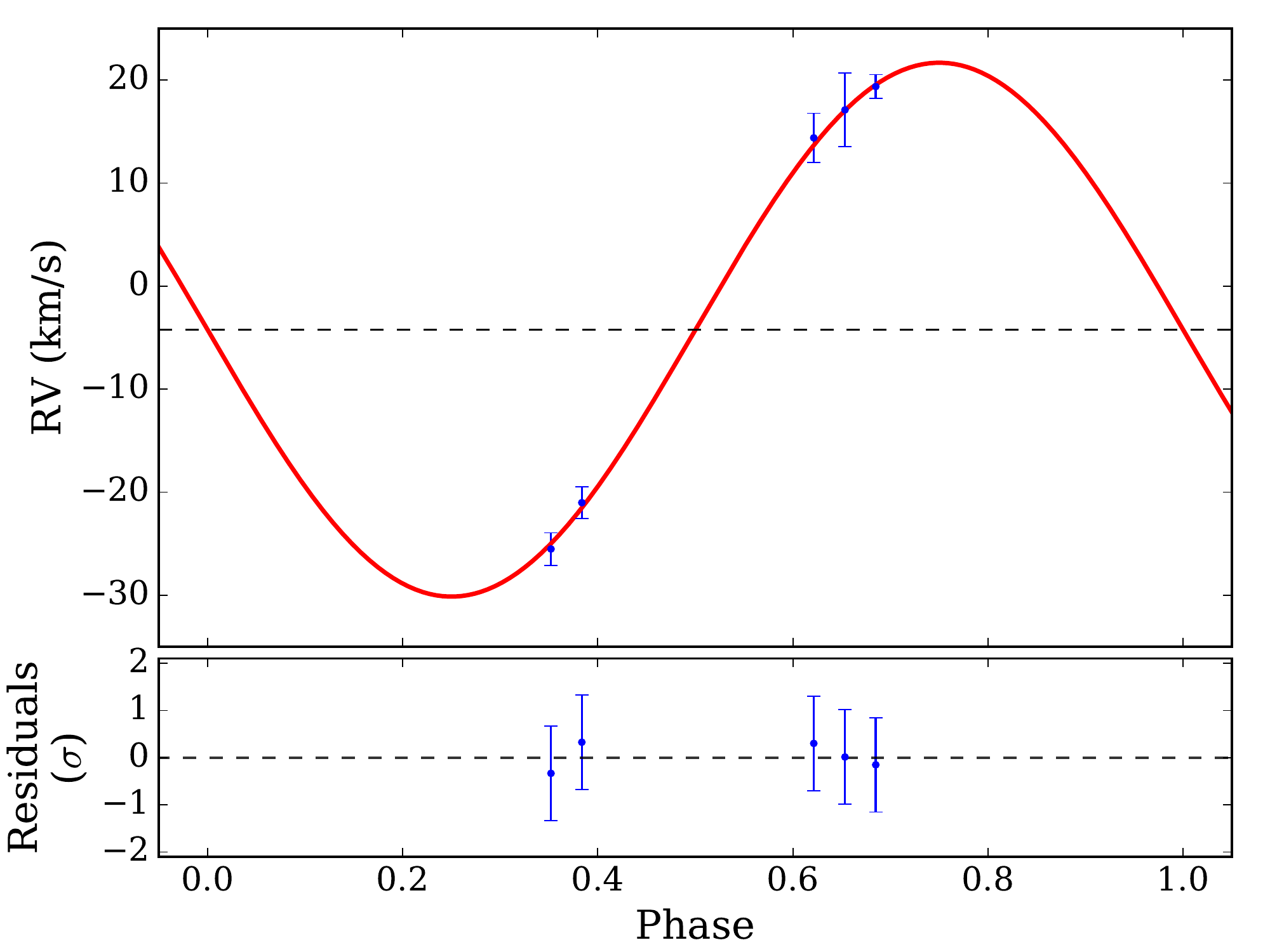}
    \caption{HARPS radial velocity data for \objname\ in blue with the best fitting radial velocity model overlaid in red. We have fixed the eccentricity of the model to 0 (as implied by the phasing of the transit and secondary eclipse in Sect.\,\ref{sec:fitting}). We also fixed the period and epoch to those measured from our transit fitting. Bottom: Residuals of the model fit.}
    \label{fig:RV_plot}
\end{figure}
When analysing the HARPS data to measure the radial velocity shifts due to \objname\ 
we used the standard HARPS data reduction software (DRS) to obtain \rewrite{our measured CCFs.} 
Initial analysis with the M2 mask showed no variation between phases in the CCFs, with a seemingly unchanging peak at \midvaluebkg. It was realised that due to the fast rotation of \starname, the spectral lines were too broadened for the M2 mask (which uses a fine grid of molecular lines), resulting in a low signal-to-noise ratio CCF. It was found that analysing with the K5 mask (which uses fewer lines and is less susceptible 
to the fast rotation) showed both a CCF peak due to the background source and a shallow wide peak due to the motion of \starname, shown in Fig.\,\ref{fig:ccf_plot}. The increased width of this peak is due to the fast rotation of \starname. We confirmed both peaks were also present when using earlier spectral type masks, albeit at a lower signal to noise. With the CCFs from the K5 mask we simultaneously fit all our HARPS CCFs with two Gaussians plus an additional linear background term. Each Gaussian is allowed to vary in amplitude and midpoint, but is required to have a constant width. To perform our simultaneous fitting we once again use an MCMC process with \emcee, with 200 walkers for 20,000 steps. We use the final 5000 steps to calculate our parameters and the results of our fitting are shown in Fig.\,\ref{fig:ccf_plot}. We folded the measured CCF peak midpoints in phase using the orbital period from Sect.\,\ref{sec:fitting} and fit a sinusoidal signal, shown in Fig.\,\ref{fig:RV_plot}. \rewrite{We also list the measured midpoints and amplitudes in Tab.\,\ref{tab:HARPS_rvs}.} As the orbit of \objname\ to have circularised (Sect.\,\ref{sec:fitting}), we fitted the RV data using a single sinusoid. We fixed the period and epoch of this sinusoid to the values measured from our transit fitting. \rewrite{With this fit we measure a systemic velocity of \midvaluepri\ and a semi-amplitude of \semiampvalue\ for \starname. We measure a systemic velocity of \midvaluebkg\ for \bkgstarname. Combining our measurement of the semi-amplitude for the radial velocity curve with the mass of \starname\ }we calculate a mass of \massvalueone\ for the transiting source for scenario (i). For scenario (ii) we obtain a value of \massvaluetwo. 

The measured \massvaluetwo\ mass for scenario (ii) places \objname\ 
within the brown dwarf regime, making the system an brown dwarf transiting a main sequence M star. Our result for scenario (i) places \objname\ at the upper end of the brown dwarf regime, near the hydrogen-burning mass limit of $\sim$ 78\mjup\ \citep[][]{Chabrier00}.  

\subsection{Rotational Broadening} \label{sec:broadening}
We can also use our HARPS data to investigate the level of rotational broadening for \starname\ \rewrite{and in turn constrain our radius measurement. We can construct a lower limit by assuming the profile of \bkgstarname\ is non-rotating and assuming a \citet{gray_2005} profile to artificially broaden it to match the profile of \starname. We have assumed a limb darkening coefficient for the rotational profile of 0.55. Artificially broadening the CCF of \bkgstarname\ gives a lower limit of 31\kms\ for \vsini, equivalent to a radius of 0.41\Rsun. This value is greater than the measured radius for \starname\ for scenario (ii) (a main sequence system with $R_{A}$=0.34\,\Rsun) and is only consistent with scenario (i) (a pre-main sequence system with $R_{A}$=0.61\,\Rsun).}

\subsection{Secondary Eclipse and Brown Dwarf temperature} \label{sec:secondary}
As part of our fitting of the NGTS lightcurve we have identified evidence of a secondary eclipse for \objname, shown in Fig.\,\ref{fig:transit_plot}. The presence of a secondary eclipse by its very nature implies non-negligible levels of flux from the brown dwarf itself. To estimate the temperature of \objname\ we equate the depth of the secondary eclipse to the ratio of fluxes in the NGTS bandpass,
\begin{equation} \label{eq:secondary}
    \delta_{eclipse} = \bigg(\frac{R_{BD}}{R_{A}}\bigg)^{2}\frac{\int F_{BD}(T_{BD})\,S(\lambda)\,d\lambda}{\int F_{A}\,S(\lambda)\,d\lambda} + A_{g}\bigg(\frac{R_{BD}}{a}\bigg)^{2}
\end{equation}
where $F_{\mathrm{BD}}(T_{\mathrm{BD}})$ and $F_{\mathrm{A}}$ are the SEDs of the brown dwarf (with temperature \tbd) and \starname\ respectively, $S(\lambda)$ is the transmission curve of the NGTS filter \citep[][]{Wheatley18} and \ag\ is the geometric albedo. For the SED of the primary star we use the results from our SED fitting. To generate the spectrum of the brown dwarf we have used the BT-Settl models \citep[][]{Allard12}, since the PHOENIX v2 models do not cover the full range of temperatures we wish to probe. 
For each model spectrum we have renormalised it to the distance of the primary star and to the expected brown dwarf radius. We opted to use these models instead of a blackbody due to the strong absorption features expected in the brown dwarf spectrum \citep[e.g.][]{Martin99}. We measured $\delta_{\mathrm{eclipse}}$ from the best fitting transit and spot model, making sure to correct for the effect of dilution in the NGTS bandpass. By including \ag\ we can also account for the effects of reflection. We have solved Eq.\,\ref{eq:secondary} in two limiting cases. These are \ag=0 (no light is reflected) and \ag=0.5.  Iterating \tbd\ between 1200 and 3500 K returns estimates of 2880 K (\ag=0.5) and 3200 K (\ag=0) for scenarios (i) and (ii). 

Comparing these temperatures to the \citet{Baraffe15} models for an isolated 75.5\,\mjup\ brown dwarf results in ages up to 80 Myr, depending on the chosen value of $A_{g}$. 
This is in agreement with our estimate of 55 Myr for the age of this system assuming our scenario (i) in which the system is located at the distance implied by the Gaia DR2 parallax of the companion NGTS-7B (Sect.\,\ref{sec:wide_binary}). In contrast, the \citet{Baraffe15} models for a 48.5\,\mjup\ brown dwarf is not able to match the measured temperature range at any age. This high temperature of the brown dwarf, heavily disfavours and effectively rules out scenario (ii), in which both M stars were assumed be on the main sequence and hence at a smaller distance. Note that in scenario (ii) the brown dwarf would have to have a mass that was well below the hydrogen burning limit (Sect.\,\ref{sec:radial_velocity}).

\subsection{Starspots} \label{sec:starspots}
\begin{figure}
	\includegraphics[width=\columnwidth]{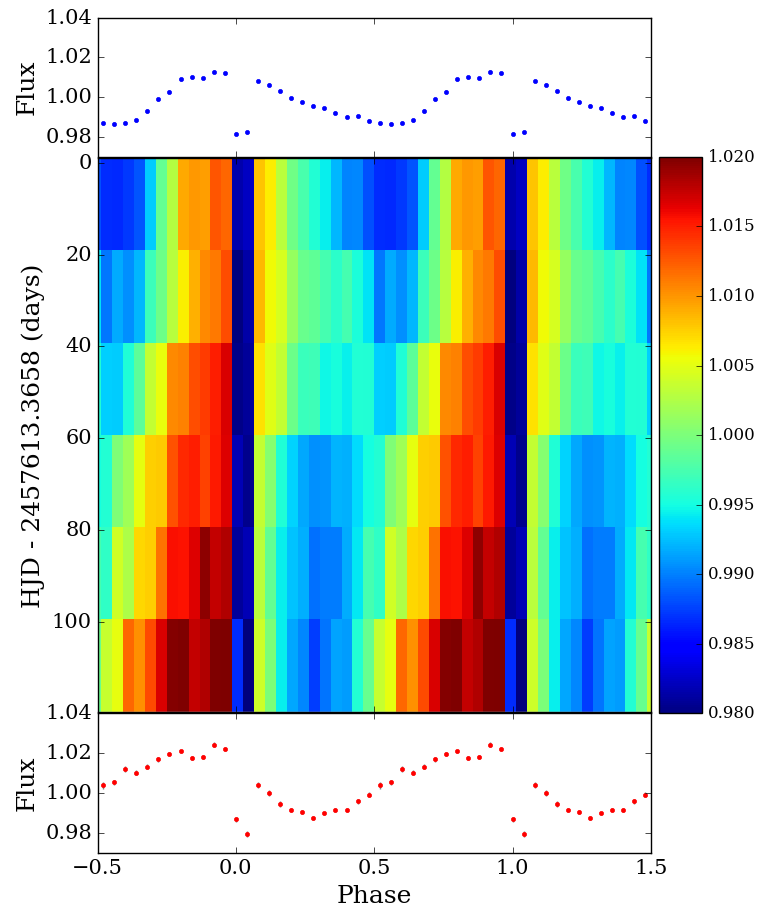}
    \caption{The evolution of starspot phase with time for \starname. Each pixel represents 0.04 in orbital phase and approximately 20 days in time. The flux is normalised to the median value of the entire NGTS dataset. The time is offset relative to 100 days into the season. Shown above and below are the phase folds corresponding to the first and last time bins respectively. The dark regions at phases 0 and 1 indicate where the transit occurs.}
    \label{fig:evolution}
\end{figure}

As part of our analysis in \rewrite{Sect.\,\ref{sec:transit_confirm} and Sect.\,\ref{sec:fitting}} we identified starspots were present in the NGTS and \TESS\ data. When fitting the NGTS data, we allowed for two starspots in our model and assumed they were representative of the average starspot behaviour of \starname. Another check for whether this modulation is due to starspots is to look for evolution throughout the NGTS lightcurve. As starspots form and dissipate they will alter both the level of lightcurve modulation and the phase at which it occurs \citep[e.g.][]{Davenport15,Jackman18a}. To search for such changes within the NGTS data we split our data into 20 day sections. Visual inspection of the phase folded lightcurve in these sections showed that the modulation was slowly changing with time, indicative of starspot evolution. To show this we have phase folded each section in bins of 0.04 in phase and plotted the flux of each phase folded lightcurve against time in Fig.\,\ref{fig:evolution}, following the method of \citet{Davenport15}.

From Fig.\,\ref{fig:evolution} we can see the movement of a dominant starspot group from around phase 0.5 to 0.25 over approximately 70 days. The level of modulation from this group is not constant, as the amplitude decreases at around 60 days in Fig.\,\ref{fig:evolution} before increasing again. One reason for this may be the dissipation and then formation of starspots from a large active region, which would act to change the overall level of modulation. From Fig.\,\ref{fig:evolution} we cannot identify any regions in the lightcurve where the starspot modulation disappears completely, meaning we are unable to measure the full starspot lifetime. Observations of M stars with \kepler\ have shown that they can have starspots with lifetimes on the order of years \citep[][]{Davenport15,Giles17}. Consequently, it is not unexpected that we do not observe drastically changing spot regions within the NGTS data alone.

A possibility for the apparent shift in starspot phase is that the stellar spin period is not exactly the same as the orbital period. A small enough offset may appear as a drift in phase without appearing as an anomaly in the phase folded data. We use Fig.\,\ref{fig:evolution} to estimate what this drift may be, by assuming that the starspot distribution remains constant and using the change in phase as an estimate of the period difference. From this we find a shift of -0.28 in phase over 100 days, approximately equal to a difference of 162 seconds per day. The starspots moving backwards in orbital phase would imply the star is spinning slightly faster than the orbital period ($P_{\mathrm{orb}}/P_{\mathrm{spin}} = 1.002$). One check for this is to mask the transits out of the original NGTS lightcurve and search for periodicity in the remaining data using a generalised Lomb-Scargle periodogram, using the {\scshape astropy} LombScargle package \citep[][]{astropy13}. Doing this and using 20,000 steps between 0 and 2 days returns a best fitting period of 16.204 hours, implying that \starname\ is slightly supersynchronous, spinning roughly 1 minute faster than the orbital period. From a sample of \kepler\ eclipsing binaries \citet{Lurie17} noted a subset of short period, slightly supersynchronous systems. It was suggested that the slight supersynchronous nature of these systems may be due to them having a non-zero eccenticity (yet too small to be measured), which may be the case for \starname.

If we assume the starspot drift is constant with time, we can calculate the expected shift during the approximately 620 day gap between the end of NGTS and the start of \TESS\ observations. We expect the starspot minimum to have shifted to phase 0.5 during the \TESS\ observations. 
However, as seen in Fig.\,\ref{fig:TESS_lc} this is where the starspot maximum occurs. This discrepancy however does not rule out the slight period difference, as the original starspot group may have decayed and been replaced by a new one at a different phase \citep[e.g.][]{Jackman18a}.

\subsection{Magnetic Activity} \label{sec:magnetic_activity}

Along with the presence of starspots, \starname\ shows other clear signatures of magnetic activity.
For instance, this source was originally highlighted as part of the NGTS flare survey. To find flares in the NGTS data, lightcurves are searched night by night for consecutive outliers about a set threshold. Full information about our detection method can be found in \citet{Jackman18a,Jackman18b}. From this process we identified four flares in the NGTS lightcurve and we have also identified one from our SAAO follow up lightcurve, which can be seen in Fig.\,\ref{fig:follow_up}. To calculate the flare energies we follow the method of \citet{Shibayama13} and have assumed the flare can be modelled as a 9000\,K blackbody. When calculating the flare energy, we have corrected each lightcurve for the expected dilution in the respective bandpass using our best SED fits from Sect.\,\ref{sec:SED}. From this we calculated energies ranging between \minflareneergyone\ and \maxflareenergyone\, or \minflareneergytwo\ and \maxflareenergytwo\ for scenarios (i) and (ii) respectively. Based on the total observing time in the NGTS and \textit{I} band filters, we measure the rate of flares above the minimum measured energy for \starname\ as \flareoccrate\ per year. The high rate of flares is similar to that of other known active M stars, such as GJ 1243 \citep[][]{Ramsay13,Hawley14} and YZ CMi \citep[][]{Lacy76}. 

In Sect.\,\ref{sec:spectrum} we noted the presence of emission lines from the Balmer series, helium and calcium, as shown in Fig.\,\ref{fig:spectrum}. By co-adding our HARPS spectra we were also able to identify the presence of $H\alpha$ emission. All of these emission lines are persistent, i.e. they appear in each individual spectrum, making us confident they are not just the product of a flare. 
\rewrite{In Sect.\,\ref{sec:transit_confirm} we attributed these strong emission lines to \starname\ and their} 
presence 
during quiescence is a clear sign that \starname\ is chromospherically active \citep[e.g.][]{Reid95,Walkowicz09}. Active M stars are known to show high energy flares more frequently than their inactive counterparts \citep[][]{Hawley14}, fitting in with our observation of multiple flares across datasets. 

For our NGTS and SAAO data we have also checked where the flares occur in starspot phase. We find that the flares occur in the NGTS data at phases 0.42, 0.43, 0.30 and 0.56. All of these phases are when the two dominant active regions are in view. From comparing to spot modulation in the \TESS\ lightcurve we also know that the flare observed in the SAAO follow up lightcurve occurred when the spots were in view, close to the spot modulation minimum. Previous studies of the flare-starspot phase relation for M stars have found flares appear to occur with a uniform distribution in starspot phase \citep[e.g.][]{Hawley14,Doyle18}. This uniform distribution has been explained as either flares occuring in small active region, which do not not cause detectable spot modulation, or flares occuring in permanently visible active regions.

Systems with known inclinations can constrain which latitudes are permanently visible, something not known for the majority of stars. As we believe \starname\ has been spun up by \objname\ and the system is not inclined relative to us, the only permanently visible active regions would be at the pole. The fact that none appear when the dominant starspots are not in view suggests the flares are associated with the starspots dominating the modulation, as opposed to a permanently visible polar region or smaller spots elsewhere. 

\subsubsection{X-ray Activity} \label{sec:xray_activity}
To determine the X-ray luminosity of \starname\ we have searched through available archival X-ray catalogues. NGTS-7 was detected during the Einstein 2 sigma survey conducted with the IPC instrument \citep[][]{Moran96}. It has an upper limit entry in the XMM upper limit server\footnote{http://xmm2.esac.esa.int/UpperLimitsServer/} (from an 8 second exposure slew observation) and was not detected in the ROSAT All-Sky Survey. For our analysis we have chosen to use the Einstein 2 sigma entry, due to it being a detection as opposed to an upper limit. Given a count excess of 8.1 counts over an exposure time of 1223 seconds, we obtain an Einstein IPC count rate of $6.6\times10^{-3}$ counts s$^{-1}$, with a signal-to-noise ratio of 2.35. We use the WebPIMMS interface\footnote{https://heasarc.gsfc.nasa.gov/cgi-bin/Tools/w3pimms/w3pimms.pl} to calculate the flux in the 0.2-12.0 keV energy range. When doing this we use a Galactic nH column density of $1.7\times10^{20}$ and an APEC optically-thin plasma model with $\log T = 6.5$. From this we estimate an unabsorbed flux of $1.66\times10^{-13}$ erg cm$^{-2}$s$^{-1}$ between 0.2 and 12.0 keV. For our two scenarios of Sect.\,\ref{sec:SED} we estimate $L_\mathrm{X}$ and $L_{\mathrm{Bol}}$ using the parameters from our best fitting SED. From this we obtain $\log L_\mathrm{X} = 29.2$ and \lxlbol= -2.54 and -2.53 respectively. While these values imply \starname\ is more X-ray active than stars which show saturated X-ray emission \citep[\lxlbol$\approx$ -3;][]{Pizzolato03,Wright18}, these values are within the scatter of the \citet{Wright11} sample. 
However, one has to take into account that \bkgstarname\ is within the Einstein IPC aperture, which has a spatial resolution of only ca.\ $1^\prime$. The detected flux may therefore stem from both stars together. If both are equally X-ray bright, this would reduce the \lxlbol\ level for \starname\ to -2.84. Another possibility is that the Einstein exposure covered a flare of one of the stars, therefore registering a higher X-ray flux level compared to the quiescent level. To check for very large flares and confirm  our choice of parameters in WebPIMMS we calculated the expected count rates in XMM and the ROSAT All-Sky Survey for comparison. In both cases we find find that the expected counts for the existing exposure times of XMM and ROSAT are below or at the respective upper limits. While this does not completely rule out a flare during the Einstein observation, it makes less likely; we are therefore confident that \starname\ is indeed an X-ray saturated star, fitting with our observations of rapid spin and magnetic activity.

\section{Discussion} \label{sec:discussion}

\subsection{The nature of \objname} \label{sec:nature}
With an orbital period of \periodhours\ hours, \objname\ is the shortest period transiting brown dwarf around a main or pre-main sequence star to date. It is also only the fifth known brown dwarf transiting an M star \citep[][]{Irwin11,Johnson11,Gillen17,Irwin18}. The host star is magnetically active, showing starspot modulation and flaring activity in both the NGTS and follow up lightcurves.

In Sect.\,\ref{sec:wide_binary} we formulated two possible scenarios for the nature of the NGTS-7 system. Scenario (i) places both stars at the distance implied by the Gaia DR2 parallax of \bkgstarname\ and results in a pre-main sequence system of roughly 55 Myr (Sect.\,\ref{sec:primary_mass}), while scenario (ii) assumes both stars are on the main sequence. These two scenarios resulted in brown dwarf masses of \massvalueone\ and \massvaluetwo\ respectively. 
In Sect.\,\ref{sec:broadening} we measured the rotational broadening of \starname\ and obtained a value of 31\kms, a value too high for a main sequence M star rotating with a period of 16.2 hours. In Sect.\,\ref{sec:secondary} we used the detection of the secondary eclipse of \objname\ to measure its temperature. We measured temperatures between 2880\,K and 3200\,K, depending on the geometric albedo of \objname. We found these measured temperatures could not be explained by a 48.5\mjup\ brown dwarf at any age, heavily disfavouring scenario (ii) once again. Based on these pieces of evidence we conclude that scenario (i) is the most likely scenario and that \objname\ is a \neighbourage\ Myr brown dwarf, transiting a tidally-locked chromospherically active pre-main sequence M dwarf in a state of spin-orbit synchronisation.

\subsection{Formation of \objname}

\begin{figure}
	\includegraphics[width=\columnwidth]{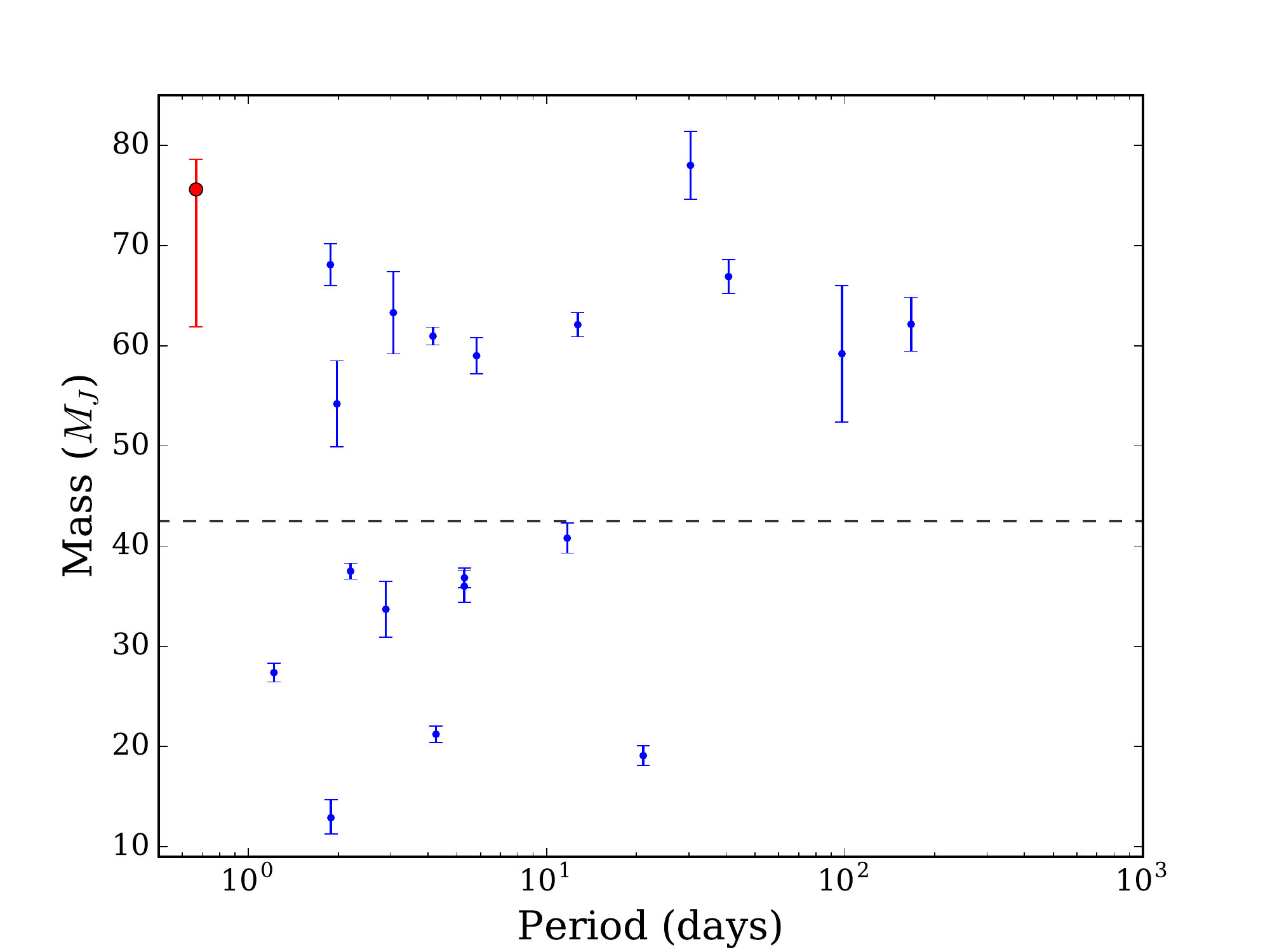}
    \caption{The mass period distribution of known transiting brown dwarfs, with the position of \objname\ from scenario (i) shown in red. This is an updated version of the same plot from \citet{Bayliss17bd}, using the table of transiting brown dwarfs compiled by \citet{Carmichael19}, along with values for AD 3116 and RIK 72 from \citet{Gillen17,David19}.
    The dashed line indicates 42.5\mjup, where \citet{Ma14} identify a gap in the mass distribution of brown dwarfs.}
    \label{fig:mass_period}
\end{figure}

It has previously been suggested that brown dwarfs around stars fall into two separate mass regimes \citep[][]{Ma14,Grieves17}, with a boundary at around 42.5\mjup. It was suggested by \citet{Ma14} that the two populations of companion brown dwarfs are related to their formation mechanism. Lower mass brown dwarfs (< 42.5\mjup) are thought to form in the protoplanetary disc, in a mechanism similar to giant planets. Whereas, higher mass brown dwarfs \rewrite{may} follow a formation path similar to stellar binaries and form through molecular cloud fragmentation. One reason for this separate mechanism is the limited mass available in protoplanetary discs to form companions, especially for discs around M dwarfs \citep[e.g.][]{Andrews13,Ansdell17}. \rewrite{Based on this analysis and the mass of \objname, we might expect molecular cloud fragmentation to be a more likely pathway for the formation of \objname.} 

If we believe that the two \Gaia\ sources are physically bound and that this is a hierarchical triple system, then \objname\ is  
similar to both NLTT41135 B \citep[][]{Irwin11} and LHS 6343C \citep[][]{Johnson11}. These systems are both M+M visual binaries where one star hosts a transiting brown dwarf. Both systems are stable with ages greater than 1 Gyr, however the presence of an outer body could help explain how \objname\ found its way onto a close orbit. \rewrite{One explanation for the tight orbit of \objname\ is that is has been moved inwards via the Kozai-Lidov mechanism \citep[][]{Kozai62,Lidov62}, where an outer body drives periodic oscillations between the inclination and eccentricity of the inner orbit. If the brown dwarf is driven into a highly eccentric orbit it may then circularise via tidal forces, resulting in both the observed tight orbit and the spin up of \starname\ \citep[e.g.][]{Bolmont12}. From a sample of 38 high mass (> 7 \mjup) exoplanets and brown dwarfs, \citet{Fontanive19} found that companions with orbital periods less than 10 days have circularisation timescales consistent with the Kozai-Lidov mechanism. We have estimated the timescale of the Kozai-Lidov mechanism ($\tau_{KL}$) for the \numbername\ system using the formalism from \citet[][]{Kiseleva98} and find $\tau_{KL}$<55 Myr for initial orbits beyond 0.1 AU. For the outer orbit (of \starname\ and \bkgstarname) we have assumed a period of 2500 years (see Sect.\,\ref{sec:orbit_of_wide_binary}) and an eccentricity of 0.5 \citep[e.g.][]{Raghavan10}. This timescale decreases for wider orbits. For the closer separations where the Kozai-Lidov mechanism may not have had enough time to operate, an alternative explanation may be that dynamical interactions during formation may have scattered \objname\ inwards and onto an eccentric orbit which was was then circularised through tidal forces.}

On the other hand, if the two \Gaia\ sources are not physically bound then \objname\ would be more similar to \gillenbd, a relatively young (sub-Gyr) brown dwarf orbiting an M star in the Praesepe open cluster \citep[][]{Gillen17}. \gillenbd\ does not show any sign of a nearby binary component and the brown dwarf is on a 1.98 day period. In this situation, the brown dwarf companion 
most likely formed close enough to its host star to migrate inwards to its current position through interactions with the primary itself \citep[e.g.][]{Armitage02}, rather than being driven to a close orbit by a third body.

As mentioned previously, one mechanism for migrating orbiting bodies inwards is through the combination of tidal forces and the magnetic wind of the host star \citep[e.g.][]{Damiani16}. These forces act in conjunction to migrate brown dwarfs inwards by transferring angular momentum from the orbit to the spin of the host star, which is then lost via magnetic breaking. The process acts with varying efficiency for different spectral types. These interactions have been argued to be particularly efficient for G and K stars \citep[][]{Guillot14}, due to their radiative interiors and moderate magnetic winds. F stars however have a much weaker wind, and the low masses and radii of M stars result in reduced tidal forces \citep[][]{Damiani16}. Both of these factors result in increased migration timescales for F and M stars. While this depends on the initial position and age of system, these interactions could provide a feasible mechanism for moving \objname\ into its current position.

\subsection{Future Evolution of \objname} \label{sec:evolution}
The remaining lifetime of \objname\ will be set by the combination of tidal dissipation and magnetic braking from the spin-down of the star which together act to remove angular momentum from the orbit of the brown dwarf. In the synchronised state, the torque on the star due to the stellar wind is equal to the tidal torque \citep[e.g.][]{Damiani15,Damiani16} and the orbit of \objname\ is expected to decay on a timescale set primarily by the magnetic braking of the host star \citep[e.g.][]{Barker09}. 

To estimate the in-spiral time $\tau_{\mathrm{a}}$ of the orbit we follow \citet{Damiani16} and use 
\begin{equation} \label{eq:tau}
    \tau_{a} \approx \frac{1}{13}\frac{h}{\alpha_{mb}C_{*}\Omega^{3}}
\end{equation}
where $h$ is the orbital angular momentum of the system, $\alpha_{\mathrm{mb}}=1.5\times10^{-14}$ is the magnetic braking parameter \citep[][]{DobbsDixon04,Damiani16}, $C_{*}$ is the primary star moment of inertia and $\Omega$ is the angular velocity of the star in the synchronised state. For our two scenarios we estimate $\tau_{\mathrm{a}}$ as 5 and 10 Myr respectively, implying that \objname\ will not remain in the current state for long and is very close to the end of its lifetime. 

This short remaining lifetime strengthens our conclusion in Sect.\,\ref{sec:nature} that NGTS-7 is a young system consisting of pre-main sequence stars and a hot brown dwarf with an age of only 55 Myr.

\subsection{The mass of NGTS-7A}\label{sec:issues}
To account for the effects of starspots on our stellar mass estimate for \starname\ in Sect.\,\ref{sec:primary_mass}, we corrected for the expected decrease in luminosity, using the age from \bkgstarname. This was then compared directly to the unspotted PARSEC models to estimate the mass. This method assumes a limiting drop in luminosity up to 10 per cent, however it may be possible that the change is greater than this. An alternative way of accounting for starspots is to use the empirical relations of \citet{Stassun12}. These relations, from observations of low mass stars and eclipsing binaries, can be used to estimate the difference between observations and models due to magnetic activity. These corrections can be utilised with either \lhlbol\ or \lxlbol. 
In Sect.\,\ref{sec:xray_activity} we estimated \lxlbol\ for the primary star, assuming both that it was the sole X-ray emitter (\lxlbol=-2.54) and that both stars were equally X-ray bright (\lxlbol=-2.84). \lxlbol\ was calculated using the best fitting SED of the primary star alone and should thus provide a more constrained estimate of the magnetic activity. Using an average of the two values with the relations for \teff\ and radius of \citet{Stassun12}, we obtain correction factors of -6.5 
per cent and 17 
per cent respectively. Applying these correction factors and comparing the new model \teff\ and radius estimates to the PARSEC models, we obtain an age of 65 Myr and a mass of 0.47\,\Msun. These values are consistent with the age and mass obtained in Sect.\,\ref{sec:primary_mass}, supporting our conclusion that magnetic activity (starspots) may have altered the SED of the primary star. 

\subsection{The orbit of the wide binary NGTS-7AB} \label{sec:orbit_of_wide_binary}
In Sect.\,\ref{sec:astrometry} and Sect.\,\ref{sec:gaia_phot} we discussed the issues present in both the \Gaia\ astrometry and photometry. Investigation of the scan angles used in \Gaia\ DR2 showed over 75 per cent of scans passed through both \starname\ and \bkgstarname. We determined that the 1.13\arcsec\ separation of the two sources was not enough to result in significant contamination of the \Gaia\ G band photometry, however would result in blended BP and RP photometry. We determined that the close proximity may be responsible for the perturbed astrometric solution of each source, due to a shifting photocentre between scans. 

Something else which has been noted as perturbing the astrometry of sources in \Gaia\ DR2 is orbital motion. \Gaia\ DR2 uses measurements obtained over an approximately two year timespan and orbital motion of a similar period could significantly affect the measured proper motions and parallaxes \citep[][]{Gaia18}. To see whether orbital motion could affect the astrometric solutions for \starname\ and \bkgstarname\ in a similar manner we have estimated the orbital period of the system, assuming a circular orbit. Using the masses of 0.48\,\Msun\ and 0.35\,\Msun\ for \starname\ and \bkgstarname\ and a separation of 173 AU we estimate the period as 2500 years. Consequently, if these sources are on a circular orbit it is unlikely orbital motion dominates the astrometric issues. We have also estimated the astrometric motion of \starname\ due to \objname\ to see whether this could be contributing to the astrometric noise. We estimate astrometric shifts of 0.012 and 0.013 mas for scenarios (i) and (ii) respectively, meaning it is unlikely \objname\ is causing significant astrometric noise \citep[see also the analysis of the GJ2069 system from][]{Mann18}.

The third data release of \Gaia\ is planned to include information about binarity \citep[e.g.][]{Lindegren18,Gaia18,GaiaHR}, meaning we will be able to constrain these scenarios further. Along with this, it is expected that blending between close sources will be improved upon. \rewrite{AO-assisted photometric observations could also help improve the} SED fitting, \rewrite{better defining the parameters of the system.}

\section{Conclusions}\label{sec:conclusions}
We have reported the discovery of \objname, a high mass transiting brown dwarf orbiting an M star with an orbital period of \periodhours\ hours. This is the shortest period transiting brown dwarf around a pre-main or main sequence star known to date and only the fifth brown dwarf transiting an M star host. Through the detection of starspot modulation in the NGTS data we have identified that the M star is in a state of spin-orbit synchronisation. We estimated an in-spiral time of 5 to 10 Myr. The short in-spiral time fits with the system being young and \starname\ being pre-main sequence M dwarf with an age of \neighbourage\ Myr. If so, then \objname\ has a mass of \massvalueone, placing it at the upper end of the brown dwarf regime. Through our analysis we identified that \starname\ is chromospherically active, showing emission lines in spectra, strong X-ray emission and exhibiting multiple flares in our photometry. These flares appear to occur more often when the starspots are in view, suggesting the two are related. 

The host star \starname\ has a neighbouring source, \bkgstarname, of similar brightness and proper motion and systemic velocity 1.13\,arcsec away. By accounting for both stars in our SED fitting, we determined the two stars to have similar temperatures. Their very similar kinematics and close proximity on the sky strongly suggest they constitute a bound binary system. If so, we believe \objname\ is part of a hierarchical triple system and the presence of \bkgstarname\ may have had a role in moving the brown dwarf into its close orbit. \rewrite{\Gaia\ DR3 and AO-assisted observations will be valuable in determining the system parameters more precisely in the future.}

\section*{Acknowledgements}
This research is based on data collected under the NGTS project at the ESO La Silla Paranal Observatory. The NGTS facility is funded by a consortium of institutes consisting of 
the University of Warwick,
the University of Leicester,
Queen's University Belfast,
the University of Geneva,
the Deutsches Zentrum f\" ur Luft- und Raumfahrt e.V. (DLR; under the `Gro\ss investition GI-NGTS'),
the University of Cambridge, together with the UK Science and Technology Facilities Council (STFC; project reference ST/M001962/1 and ST/S002642/1). 
JAGJ is supported by STFC PhD studentship 1763096 and would like to thank coffee.
PJW, SG, TL, BTG, DP and RGW  are supported by STFC consolidated grant ST/P000495/1.
SLC is supported by an STFC Ernest Rutherford fellowship.
MNG acknowledges support from MIT's Kavli Institute as a Torres postdoctoral fellow.
JSJ acknowledges support by Fondecyt grant 1161218 and partial support by CATA-Basal (PB06, CONICYT).
CAW acknowledges support from Science and Technology Facilities Council grant ST/P000312/1.
EG gratefully acknowledges support from the David and Claudia Harding Foundation in the form of a Winton Exoplanet Fellowship.

This publication makes use of data products from the Two Micron All Sky Survey, which is a joint project of the University of Massachusetts and the Infrared Processing and Analysis Center/California Institute of Technology, funded by the National Aeronautics and Space Administration and the National Science Foundation.
This publication makes use of data products from the Wide-field Infrared Survey Explorer, which is a joint project of the University of California, Los Angeles, and the Jet Propulsion Laboratory/California Institute of Technology, funded by the National Aeronautics and Space Administration.
This work has made use of data from the European Space Agency (ESA) mission
{\it Gaia} (\url{https://www.cosmos.esa.int/gaia}), processed by the {\it Gaia}
Data Processing and Analysis Consortium (DPAC,
\url{https://www.cosmos.esa.int/web/gaia/dpac/consortium}). Funding for the DPAC
has been provided by national institutions, in particular the institutions
participating in the {\it Gaia} Multilateral Agreement. This paper uses observations made at the South African Astronomical Observatory (SAAO). 




\bibliographystyle{mnras}
\bibliography{references} 


\appendix
\subsection{Affiliations}
$^{1}$Dept. of Physics, University of Warwick, Gibbet Hill Road, Coventry CV4 7AL, UK \\
$^{2}$Centre for Exoplanets and Habitability, University of Warwick, Gibbet Hill Road, Coventry CV4 7AL, UK\\
$^{3}$Institute of Astronomy, University of Cambridge, Madingley Rise, Cambridge CB3 0HA, UK\\
$^{4}$Department of Physics and Astronomy, University of Leicester, University Road, Leicester, LE1 7RH\\
$^{5}$Astrophysics Group, Cavendish Laboratory, J.J. Thomson Avenue, Cambridge CB3 0HE, UK\\
$^{6}$Department of Physics, and Kavli Institute for Astrophysics and Space Research,\\ Massachusetts Institute of Technology, Cambridge, MA 02139, USA\\
$^{7}$Juan Carlos Torres Fellow\\
$^{8}$Geneva Observatory, University of Geneva, Chemin des Mailettes 51, 1290 Versoix, Switzerland\\
$^{9}$Institute of Planetary Research, German Aerospace Center, Rutherfordstrasse 2, 12489 Berlin, Germany\\
$^{10}$Center for Astronomy and Astrophysics, TU Berlin, Hardenbergstr. 36, D-10623 Berlin, Germany\\
$^{11}$Astrophysics Research Centre, Queen's University Belfast, 1 University Road, Belfast BT7 1NN, UK\\
$^{12}$Departamento de Astronom\'ia, Universidad de Chile, Casilla 36-D, Santiago, Chile\\
$^{13}$Centro de Astrof\'isica y Tecnolog\'ias Afines (CATA), Casilla 36-D, Santiago, Chile\\
$^{14}$Instituto de Astronom\'ia, Universidad Cat\'olica del Norte, Angamos 0610, 1270709, Antofagasta, Chile\\
$^{15}$Leibniz Institute for Astrophysics Potsdam (AIP), An der Sternwarte 16, 14482 Potsdam, Germany\\
$^{16}$Institute of Geological Sciences, FU Berlin, Malteserstr. 74-100, D-12249 Berlin, Germany\\
$\ddagger$ Winton Fellow

\section{SED fitting corner plot}
\begin{figure*}
	\includegraphics[width=\textwidth]{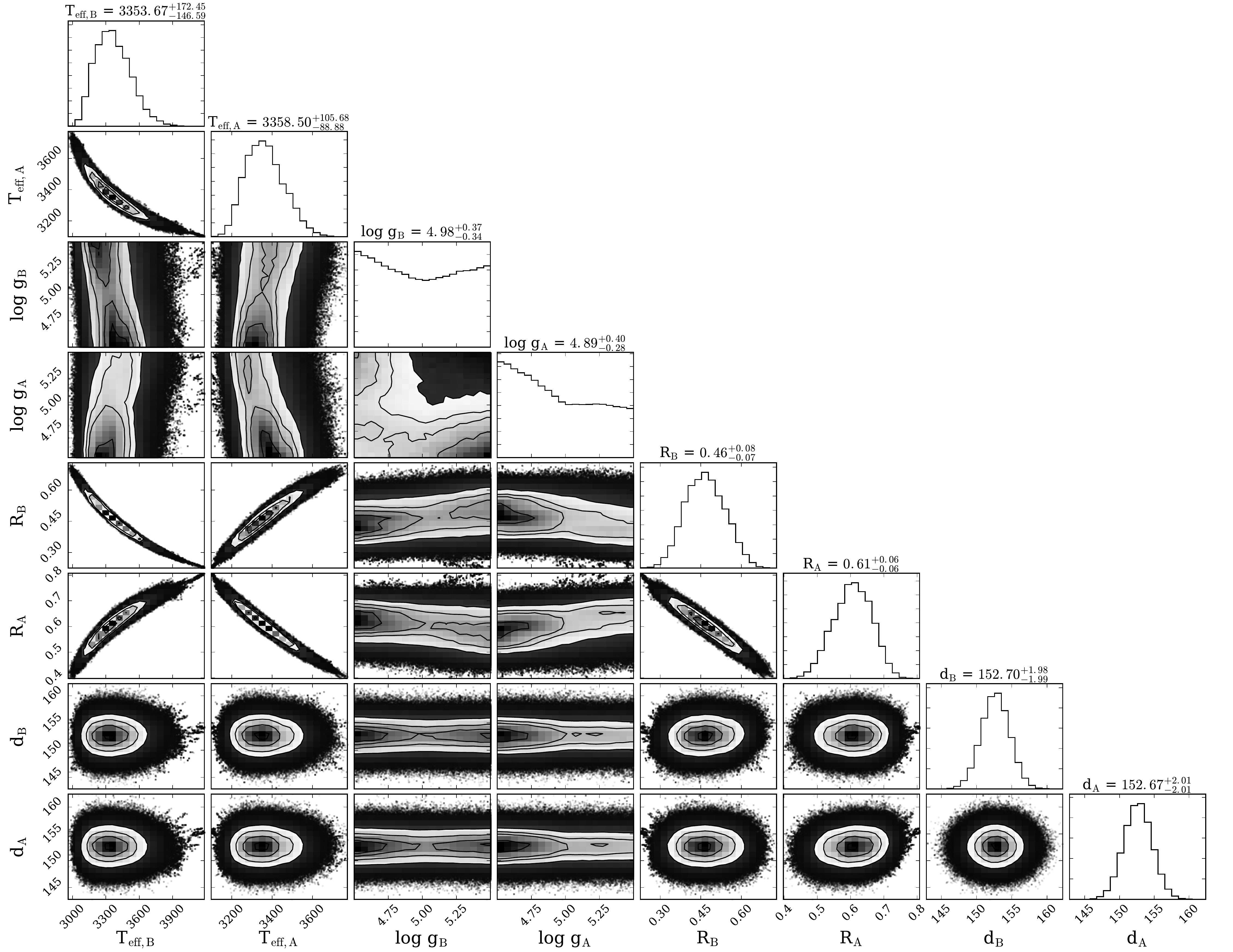}
    \caption{The full corner plot for the scenario (i) SED fitting. Note the strong correlation between \teff\ and radius, which we have accounted for in our analysis.}
    \label{fig:corner_fig}
\end{figure*}

\begin{figure*}
	\includegraphics[width=\textwidth]{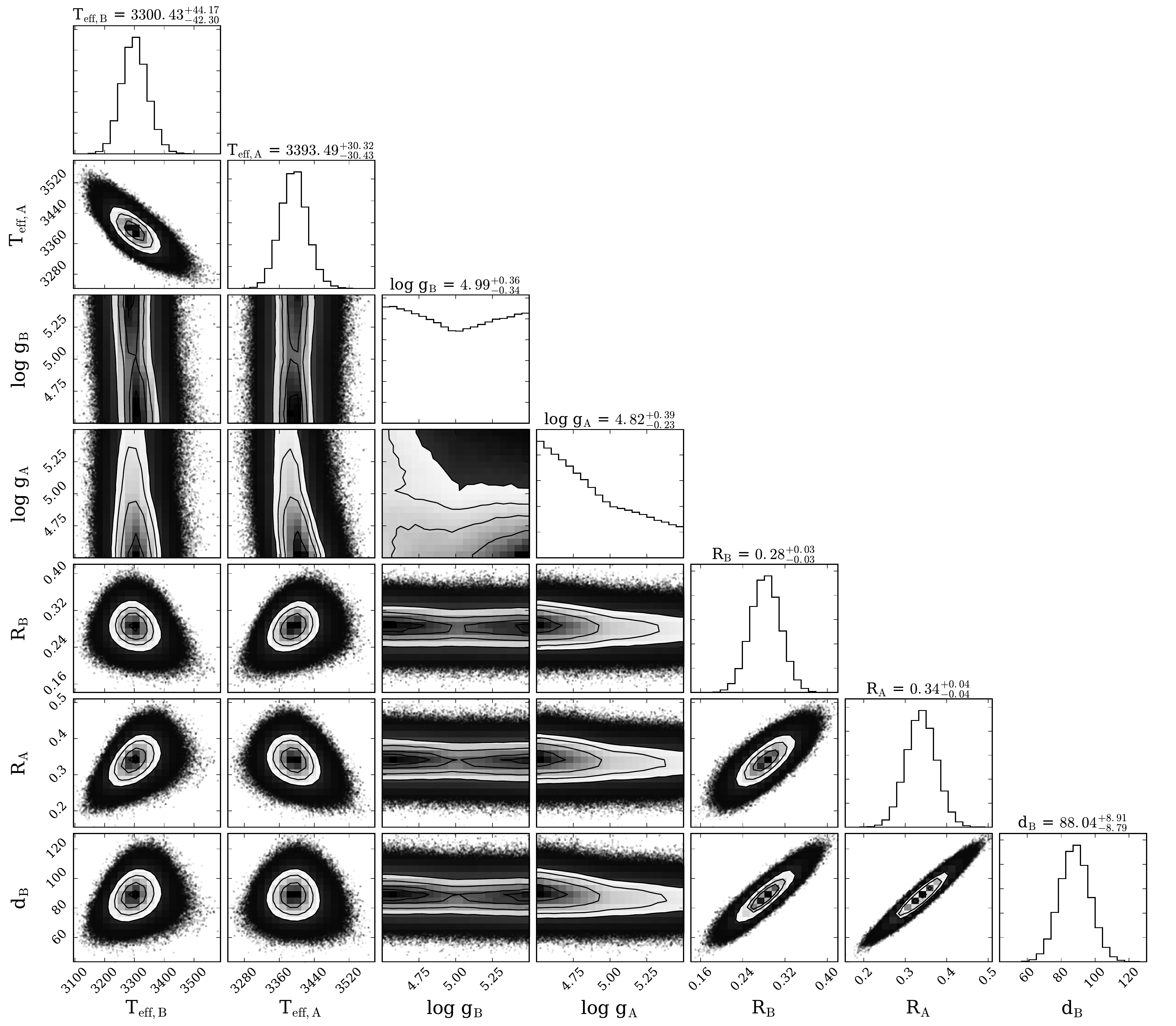}
	\caption{The full corner plot for the scenario (ii) SED fitting.}
    \label{fig:corner_fig_main_seq}
\end{figure*}


\bsp	
\label{lastpage}
\end{document}